\documentclass{aa}

\usepackage{color}
\usepackage{natbib}
\usepackage{graphicx}
\usepackage{caption}
\usepackage{subcaption}
\usepackage{txfonts}

\begin{document}

   \title{Stretch to stretch, dust to dust: lower-value local $H_{0}$ measurement from two-population modelling of type Ia supernovae}
   \titlerunning{Stretch to stretch, dust to dust: towards unbiased $H_{0}$ measurement}
   
   \author{Rados{\l}aw Wojtak \inst{1}
          \and Jens Hjorth \inst{1}
          }

   \institute{DARK, Niels Bohr Institute, University of Copenhagen, Jagtvej 155, 2200 Copenhagen, Denmark \\
              \email{radek.wojtak@nbi.ku.dk}
              }

   \date{}

  \abstract
 {}
   {
   We revisit the local Hubble constant measurement from type Ia supernovae calibrated with Cepheids (SH0ES) by remodelling the supernova data using two 
   supernova populations emerging from the observed bimodal distribution of the SALT2 stretch parameter. Our analysis accounts for population differences in both intrinsic properties (related to possible initial conditions, including supernova progenitor channels) and host-galaxy extinction (expected from well-known environmental differences associated observationally with the two populations).
   }
   {Based on a two-population Bayesian hierarchical modelling of the SALT2 light-curve parameters from the Pantheon+ compilation, we simultaneously constrain intrinsic and extrinsic properties of both supernova populations, match probabilistically the calibration supernovae with the corresponding population in the Hubble flow, and derive the Hubble constant.
   }
   {
   The difference between the two supernova populations is primarily driven by their mean absolute magnitudes and total-to-selective extinction coefficients. This is related but not equivalent to the traditional mass-step correction (including its broadening for reddened supernovae). The mean extinction coefficient of the supernova population used to propagate distances from the calibration galaxies to the Hubble flow is found to be consistent with the Milky Way-like interstellar dust model with $R_{\rm B}\approx 4$ and substantially higher than the extinction model assumed in the SH0ES measurement. Allowing for possible differences between reddening in the calibration galaxies and the corresponding population in the Hubble flow, we obtain $H_{0}=70.59\pm{1.15}$~km~s$^{-1}$~Mpc$^{-1}$. For the most conservative choice assuming equal prior distributions, we find $H_{0}=71.45\pm{1.03}$~km~s$^{-1}$~Mpc$^{-1}$.
   }
   {Our reanalysis of type Ia supernovae results in a reduction of the discrepancy with the CMB-based Planck value of $H_{0}$ by at least 30 per cent ($3.5\sigma$) and up to 50 per cent ($2.2\sigma$). We discuss the correspondence between our result and similar low-value estimates previously obtained from about 10 nearest calibration galaxies.
    }


   \maketitle

\section{Introduction}

The Hubble constant measurement from the Supernova H0 for the Equation of State (SH0ES) programme \citep{Riess2022} is 
about $5\sigma$ discrepant with the CMB-based measurement from Planck observations assuming a flat $\Lambda$CDM cosmological 
model. Complementary CMB constraints \citep[e.g. lensing power spectrum,][]{Qu2025}, measurements from independent 
instruments \citep[e.g. the South Pole Telescope,][]{Balkenhol2023} or fitting the Planck data with alternative likelihood models 
\citep{Efstathiou2024} yield results which are consistent with the Planck cosmological model. This leaves very little room either for 
hidden systematic effects in the CMB observations or for pre-recombination modifications of the standard model devised to increase the Hubble constant derived from the CMB.

A growing number of independent studies point to possible underestimated systematic uncertainties in the SH0ES measurement. 
The main focus has been put on observational tests or alternative modelling of the Cepheid data sector 
\citep[see e.g.][]{Freedman2024,Kushnir2025,Hogras2025}. However, particularly concerning although not fully appreciated are 
intrinsic anomalies found in the supernova data and  associated with an overestimation of the Hubble constant 
\citep{Wojtak2022,Perivolaropoulos2023,Wojtak2024,Gall2024,Hoyt2025}. These anomalies can be traced back, either directly 
\citep{Wojtak2024} or indirectly (see section~\ref{sec:summary} and discussion therein), to the unresolved problem of how to model environmental effects on type Ia supernova brightness in a way that is both accurate and consistent with basic astrophysical properties such as interstellar extinction expected 
in supernova host galaxies.

It has long been known that type Ia supernova properties vary across
host-galaxy types and local stellar environments. Following early evidence for a relation between the width (decline rate) of type Ia supernova light curves \citep[driven by the mass of radioactive $^{56}$Ni;][]{Arnett1982} and host-galaxy stellar populations \citep{Hamuy2000}, observations revealed tight correlations between this light curve parameter and the host-galaxy stellar age \citep{Sullivan2006}, luminosity-weighted age of host galaxy \citep{Howell2009} or the local specific star formation rate \citep{Rigault2013,Rigault2020}. Host galaxies modulate also supernova 
peak magnitudes, with variations which are not captured by first order 
standardisation methods such as the Tripp calibration \citep{Tripp1998}. 
The relation is commonly studied as a (non-linear) correlation
between supernova peak magnitude and the host-galaxy stellar mass \citep{Kelly2010,Sullivan2010,Smith2020}. Similar or statistically more 
significant correlations were also shown for other host-galaxy properties, for example the local specific star formation rate \citep{Rigault2020}. The physical nature of this multifaceted relation between type Ia supernova properties and their environment is not fully understood. Some models attempt 
to reduce the problem solely to the effect of extinction and its dependence on host galaxy properties \citep{Brout2021,Popovic2021}. 
However, independent analyses of type Ia supernova data \citep[see e.g.][]{Wojtak2023,Gonzalez2021,Grayling2024}, tests against host galaxy properties beyond the stellar mass \citep{Duarte2023,Kelsey2023}, simulations of extinction in type Ia supernova sight lines \citep{Hallgren2025}, and modelling the role of the progenitor age \citep{Wiseman2022,Wiseman2023} point to a relevant contribution from supernova intrinsic properties correlated with the host-galaxy environment.

Due to the fact that Cepheids are observed only in young late-type galaxies, the SH0ES programme introduces a strong implicit selection 
bias in type Ia supernova environments. With type Ia supernovae from the Hubble flow found in galaxies of all morphological types, the 
calibration galaxies (those with observed Cepheids and type Ia supernovae) do not constitute a representative sample of the Hubble flow. 
The SH0ES team attempts to mitigate the selection bias by restricting the Hubble flow to galaxies resembling those in the calibration sample 
\citep{Riess2022}. However, their extinction model is trained on the unrestricted Hubble flow data and then extrapolated to the calibration galaxies \citep{Popovic2021,Brout2022}.

The extinction model employed in the Pantheon+ supernova compilation \citep{Brout2022} and subsequently used by \citep{Riess2022} to measure 
the Hubble constant ascribes different total-to-selective extinction coefficients to host galaxies with stellar mass larger or smaller than $10^{10}M_{\odot}$ 
\citep{Brout2021}. The B-band extinction coefficient measured from high stellar-mass host galaxies in the cosmological sample is $R_{\rm B}\approx 3$ \citep{Popovic2021}. 
Despite the fact that the high-mass bin contains both early- and late-type host galaxies with comparable fractions, the model extrapolates the same extinction coefficient to 
the corresponding high stellar-mass calibration galaxies, 
which happen to make up 70 per cent of the entire calibration sample. The extrapolated extinction coefficient agrees neither with the most 
likely value of $R_{\rm B}\approx 4$ expected in the calibration galaxies as close analogs of the Milky Way \citep[with average $R_{\rm B}$ ranging between 4.1 and 4.3][]{Fitzpatrick2007,Schlafly2016} nor with the extinction curve employed by \citet{Riess2022} for extinction corrections in Cepheids ($R_{\rm B}=4.3$) nor with recent extinction measurement for NGC 5584, a SH0ES calibration galaxy \citep{Murakami2025}. 
It is therefore not surprising that this approach results in an intrinsic tension with distances measured from Cepheids, where the 12 reddest ($c>0$) supernovae with $R_{\rm B}\approx 3$ 
(29 per cent of all calibrators) are $2.3\sigma$ fainter relative to random subsamples of equal size \citep{Wojtak2024}. 
This tension can be eliminated by assuming consistently the same Milky Way-like extinction coefficient in all calibration galaxies. 
The correction improves the fit quality in the calibration data block and results  in a decrease of the derived Hubble constant by $1.7$~km~s$^{-1}$~Mpc$^{-1}$ 
(30 per cent of the SH0ES-Planck difference) or $2.9$~km~s$^{-1}$~Mpc$^{-1}$ (50 per cent of the SH0ES-Planck difference) for an independently modelled 
distribution of reddening in the calibration sample \citep{Wojtak2024}. Comparable corrections were obtained by \citet{Rigault2015} from modelling an empirical 
relation between type Ia supernova brightness and host-galaxy star formation rate, and quantifying the impact of the apparent difference between the calibration sample and the Hubble flow on the Hubble constant determination, 
based on early SH0ES data from \citet{Riess2011}. These studies show that the environmental effects are likely underestimated in the SH0ES measurement of the Hubble constant and give rise to a bias which may account for up to 50 per cent of the difference between the SH0ES and Planck values of the Hubble constant.

Controlling the effect of supernova environment on distance propagation is key for obtaining unbiased measurements of the Hubble constant. Sufficiently informative 
differentiation between local supernova environments can be realised using only supernova light curve parameters. 
The primary motivation lies in the observation that the decline rate of type Ia supernova light curves, as measured by the stretch parameter of the SALT light curve fitter 
\citep{Guy2007}, correlates well with the local supernova environment quantified by the specific star formation rate \citep{Rigault2020}. Furthermore, the observed distribution of stretch parameter exhibits bimodality which is well visible in supernova compilations \citep{Scolnic2018} and recently obtained volume-limited sample from the Zwicky Transient Factory \citep{Ginolin2025}. The bimodality is thought to arise from incomplete mixing of two supernova populations associated with high- and low-star formation environments \citep{Rigault2020}, and perhaps correlated with single and double-degenerate progenitor channels, respectively \citep{Nicolas2021}.

We propose to use the stretch parameter and the bimodal feature in its distribution to match supernova environments between the calibration sample and the Hubble flow, as a data-processing step integrated with standardisation modelling. The match is done in a probabilistic way by using recently developed two-population Bayesian hierarchical model \citep{Wojtak2023}. The model allows to separate probabilistically high-stretch and low-stretch supernova 
populations, and constrain supernova and extinction properties separately in the two populations. Environment-related bias in the Hubble constant determination is expected to be minimised thanks to the fact that distance measurements are propagated between the calibration 
sample and the Hubble flow using the same supernova population associated with the high-stretch peak of the stretch distribution. Fitting the model to supernova data yields also a range of constraints on supernova intrinsic and extrinsic properties. Initial analyses of supernova data in the Hubble flow show that the high-stretch supernova population 
is consistent with a Milky Way-like extinction with $R_{\rm B}\approx 4$ \citep{Wojtak2023}. Our goal, therefore, is not only to revise the Hubble constant measurement but to develop 
a standardisation model which can be reconciled with a Milky Way-like extinction expected in late-type galaxies like those in the calibration sample. In order to isolate the impact of modelling type Ia supernova on the Hubble constant derived from the SH0ES and Pantheon+ data, we keep the distance calibrations (best-fit distance moduli and the full covariance matrix) derived by \citet{Riess2022} from Cepheids unchanged.

The outline of the paper is as follows. In section~\ref{sec:data}, we describe the supernova data from the Pantheon+ compilation. In section~\ref{sec:model}, we outline the two-population model, its observational motivation and assumptions, and observationally motivated difference between the Hubble flow and the calibration sample in terms of population mixing. We also discuss the likelihood function for fitting the Hubble flow and the calibration data. Section~\ref{sec:results} shows the results of data modelling including constraints on intrinsic and extrinsic properties of supernovae from the two populations and derived values of the Hubble constant. We summarise and discuss the obtained results in section~\ref{sec:summary}.

\section{Data}
\label{sec:data}

We use light curve parameters of type Ia supernovae from the Pantheon+ compilation \citep{Scolnic2022_lc}. The parameters were obtained 
from fits based on the SALT2 model developed by \citet{Guy2007} and retrained by \citet{Taylor2021}. They include the apparent rest-frame $B$-band magnitude 
$m_{\rm B}$, the dimensionless stretch parameter $x_{1}$ and the colour parameter $c$ which closely approximates the rest-frame $B-V$ colour at rest-frame 
$B$-band peak \citep{Brout2021}. Our analysis includes light curve measurements given by best-fit parameters and covariance matrices obtained for 
each individual supernova.

Our approach involves modelling the distribution of type Ia supernovae in the light curve-parameter space. This requires that every distinct supernova 
is represented by a single measurement. In order to comply with this requirement, we combine all duplicates (light curve parameters 
from different surveys included in Pantheon+) of each distinct supernova into single measurement. We assume that the combined measurements are given by the product 
of Gaussian probability distributions with the mean values and covariance matrices given by duplicates of the same distinct supernova. We also account for apparent 
discrepancies between duplicates by fitting an extra scatter in each of the light curve parameters. The scatter is modelled as three independent parameters 
coadded to the diagonal elements of duplicates' covariance matrices. This approach yields more accurate best-fit values and more realistic errors 
(covariance matrices) than combining Gaussian distributions with the original covariance matrices. The extra scatter in $\{m_{\rm B},x_{1},c\}$ per duplicate is on average $\{12,90,13\}$ per cent of the errors from the catalogue for the calibration sample, and $\{19,93,13\}$ per cent for duplicates in the Hubble flow.

For the Hubble flow, we select all supernovae in the redshift range between $0.015$ and $0.15$ (536 supernovae, excluding SN 2007A which is a calibrator). 
Including more low-redshift supernovae ($0.015<z<0.023$) than \citet{Riess2022} has negligible impact on the Hubble constant estimation 
($\Delta H_{0}\lesssim 0.1$~km~s$^{-1}$~Mpc$^{-1}$, based on Pantheon+ likelihood) but it results in more constraining power 
for two-population modelling, from the distribution of $c$ and $x_{1}$. The light curve parameters and supernova Hubble residuals with respect to the 
standard Tripp calibration are shown in Appendix \ref{sec:appA} (Figure~\ref{data_HF_app}). Figure~\ref{data_HF_app} also shows 5 SNe 
(2010ai, 2009dc, 2007ci, 1998ab, 2006cz) which are identified as $\geq 3\sigma$ outliers with respect to the best fit two-population model 
(with about 1 supernova expected at this probability threshold). In order to minimise potential impact of these supernovae on the estimations of some 
model parameters, they are omitted from our analysis.

We calibrate type Ia supernova intrinsic brightness using distance moduli of 37 calibration galaxies obtained from Cepheid observation of the SH0ES programme. 
Specifically, we use best-fit distance moduli and uncertainties derived exclusively from Cepheid data (without inclusion of any supernova in any host) and provided in Table~2 of \citet{Riess2022}. We account for correlations between the errors by employing the full covariance matrix. We compute its non-diagonal elements 
from a Markov chain generated with the publicly available code from the SH0ES collaboration\footnote{https://github.com/PantheonPlusSH0ES}. The supernova 
data block of the likelihood was disabled so that the posterior probability sampling was run based on the Cepheid data and only with Cepheid parameters 
(including 37 distance moduli of the calibration galaxies).

\section{Model}
\label{sec:model}

We adopt a commonly used model of relations between the SALT2 type Ia supernova light curve parameters $\pmb{\xi}=\{m_{\rm B},x_{1},c\}$ 
(observables) and a minimum range of physically motivated latent variables. The model is given by the following set of equations
\begin{equation}
\begin{aligned}
m_{\rm B} & =M_{\rm B}+\mu-\alpha X_{1} +\beta c_{\rm int}+R_{\rm B}E(B-V) \\
x_{1} &=X_{1} \\
c&=c_{\rm int}+E(B-V),
\end{aligned}
\label{observables}
\end{equation}
where $c_{\rm int}$ is the supernova intrinsic colour (at the peak of the $B$-band light curve), $E(B-V)$ is the host-galaxy 
reddening, $R_{\rm B}$ is the total-to-selective extinction coefficient, $M_{\rm B}$ is the supernova absolute magnitude, $\alpha$ and 
$\beta$ are free coefficients and $\mu$ is the distance modulus. We formally distinguish between the observed stretch parameter $x_{1}$, 
which is subject to a perturbation from measurement errors, from the latent stretch parameter $X_{1}$, whose distribution is driven 
by its prior distribution. With this distinction, we can write our model in a compact form as $\pmb{\xi}=\pmb{\xi}(\pmb{\phi})$, where 
$\pmb{\phi}=\{M_{\rm B},X_{1},c_{\rm int},E(B-V),\alpha,\beta,R_{\rm B},\mu \}$ is the vector of all latent variables.

\subsection{Likelihood and priors}
We employ the likelihood which models the observed distribution of type Ia supernovae in the light curve-parameter space, in relation to the 
measurements errors and the prior distribution of the latent variables. The likelihood can be expressed in the following way \citep[see e.g.][]{Mandel2017,Wojtak2023}
\begin{equation}
L \propto \prod_{i}^{N} p(\pmb{\xi_{i}}|\pmb{\Theta})=
\prod_{i}^{N}\int \mathcal{G}[\pmb{\xi}(\pmb{\phi});\pmb{\xi_{\rm obs\,i}},\mathsf{C_{\rm obs\,i}}]
p_{\rm prior}(\pmb{\phi}|\pmb{\Theta})\textrm{d}\pmb{\phi},
\label{likelihood}
\end{equation}
where $\Theta$ is the vector of all hyperparamers parametrising the prior distribution $p_{\rm prior}$, $\pmb{\xi_{\rm obs\,i}}$ is the vector 
of the measured (best-fit) light curve parameters ($i$-th supernova), $\mathsf{C_{\rm obs\,i}}$ is the corresponding covariance matrix ($i$-th 
supernova). $\mathcal{G}(\pmb{x};\pmb{\mu_{x}},\mathsf{C_{x}})$ denotes a Gaussian distribution with mean $\pmb{\mu_{x}}$ and covariance matrix 
$\mathsf{C_{x}}$ and the product in eq.~\ref{likelihood} runs over all independent type Ia supernovae.

We assume Gaussian prior distributions for $\{M_{\rm B},c_{\rm int},X_{1},R_{\rm B}\}$ and single-value distributions (the Dirac delta function) 
for ${\alpha,\beta}$. We will hereafter adopt the notation in which $\widehat{\phi}$ and $\sigma_{\phi}$ denote the mean and the dispersion 
of the prior Gaussian distribution of variable $\phi$. Integration over these latent variables results in a 
Gaussian distribution with mean and covariance matrix given by the corresponding hyperparameters and the remaining latent variables \citep[$\mu$ and host-galaxy reddening; see Appendix of][]{Wojtak2023}. The remaining integration over $E(B-V)$ is performed numerically. Here, we adopt the prior distribution given by the exponential model, i.e.
\begin{equation}
p(E(B-V))=\frac{1}{\tau}\exp\Big(-\frac{E(B-V)}{\tau}\Big),
\end{equation}
where $\tau$ is the mean reddening $\langle E(B-V)\rangle$ (unless restricted supernova colours are considered). The exponential distribution provides a good approximation to the interstellar reddening expected in late-type host galaxies under the assumption that type Ia supernovae 
trace the stellar component, but it does not reproduce the simulated interstellar reddening in early-type host galaxies \citep{Hallgren2025}. Although possible alternatives 
to the exponential model are not ruled out and there are ongoing studies testing them against observations \citep{Wojtak2023,Ward2023}, the exponential distribution 
remains a benchmark model for the prior host-galaxy reddening distribution and as such it is also implemented in our analysis. 

The adopted priors are also motivated by the observed trend of intrinsic scatter in supernova Hubble residuals increasing with colour \citep[see e.g. Figure 2 of][]{Brout2021}. 
Intrinsic scatter is mapped in our model model primarily onto a combination of scatter in $M_{\rm B}$ (achromatic component) and $R_{\rm B}$ (scatter component increasing 
with supernova colour). Although some previous studies attempted to account also for a possible scatter in $\beta$, observational constraints on this component are not significantly different from 0 \citep{Brout2021}.

Distance moduli of the supernovae in the Hubble flow depend on cosmological model. We assume the Planck cosmological model \citep{Planck2020_cosmo} for the 
shape of $\mu(z)$ and free Hubble constant $H_{0}$, i.e.
\begin{equation}
\mu(z)=\mu_{\rm Planck}(z)-5\log_{10}(H_{0}/H_{0,{\rm Planck}}).
\end{equation}
We account for the main uncertainty related to the peculiar velocity by including the following error in distance modulus to each supernova
\begin{equation}
\sigma_{\mu}(z)=\frac{5}{\ln10}\frac{\sigma_{\rm v}}{cz},
\end{equation}
where $\sigma_{\rm v}=250$km~s$^{-1}$. We assume vanishing correlations between different supernovae in the Hubble flow so that integrating over $\mu$ 
in eq.~(\ref{likelihood}) results effectively in adding $\sigma_{\mu}(z)$ in quadrature to $\sigma_{M_{\rm B}}$. Due to correlations between distance measurements 
from Cepheid observations (related to shared distance calibration and supernova siblings present in 4 calibration galaxies), 37 distance moduli of the calibration galaxies are treated as nuisance parameters in all fits involving the calibration data. In these cases, the posterior probability distribution sampled by means of Markov Chain Monte Carlo (MCMC) method includes 
the Gaussian prior given by the best-fit distance moduli to the calibration galaxies and the corresponding covariance matrix obtained by \citet{Riess2022}.

The Pantheon+ catalogue discards supernovae with stretch parameters $|x_{1}|>3$ or colours $|c|>0.3$. However, as it can be seen in Figure~\ref{data_HF_app}, these cuts have 
negligible impact on sampling the true distribution of supernovae in the Hubble flow. Therefore, without a loss of precision we normalise the probability density 
$p(\pmb{\xi_{i}}|\pmb{\Theta})$ in eq.~(\ref{likelihood}) of the Hubble flow data block without using any bounds on light curve parameters. However, this normalisation is not valid for the calibration sample where supernovae were selected using a narrower range of the colour 
parameter \citep{Riess2022}. The colour cut adopted as a selection criterion discard any supernovae with $|c|>0.15$ and has a strong impact on the observed 
colour distribution. We account for this observational selection effect by normalising the probability density $p(\pmb{\xi_{i}}|\pmb{\Theta})$ in the range $|c|<0.15$ 
(observed colours). The adopted normalisation means that our modelling employs a selection function in the form of a top-hat filter in the observed supernova colour. 
Incorporating the selection function in the probability model mitigates potential biases in the dust parameter estimation in the calibration sample. 
Due to a correlation between supernova stretch and host-galaxy environment \citep{Nicolas2021,Rigault2020}, the stretch parameter distribution of the calibration 
supernovae is primarily influenced by the effect of selecting exclusively late-type host galaxies with observable Cepheids. As shown on the right panel of 
Figure~\ref{x1_hist}, the selection effect strongly suppresses the stretch distribution at $x_{1}<-0.5$ but it keeps the high-stretch tail practically unchanged. 

\subsection{Two populations}
\label{sect:twopop}

\begin{figure*}
    \centering
    \begin{subfigure}[l]{0.45\textwidth}
        \centering
        \includegraphics[width=\linewidth]{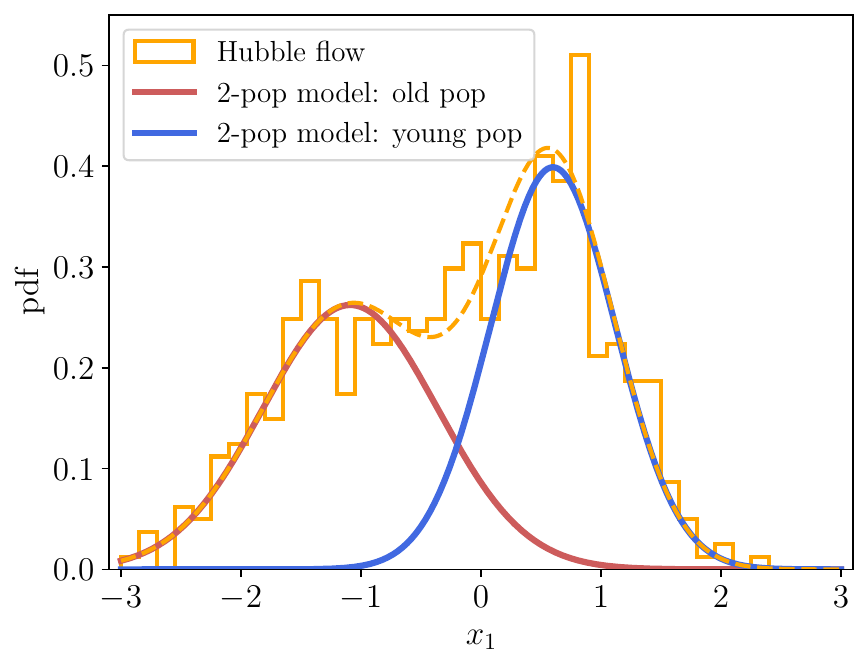} 
    \end{subfigure}
    \begin{subfigure}[r]{0.45\textwidth}
        \centering
        \includegraphics[width=\linewidth]{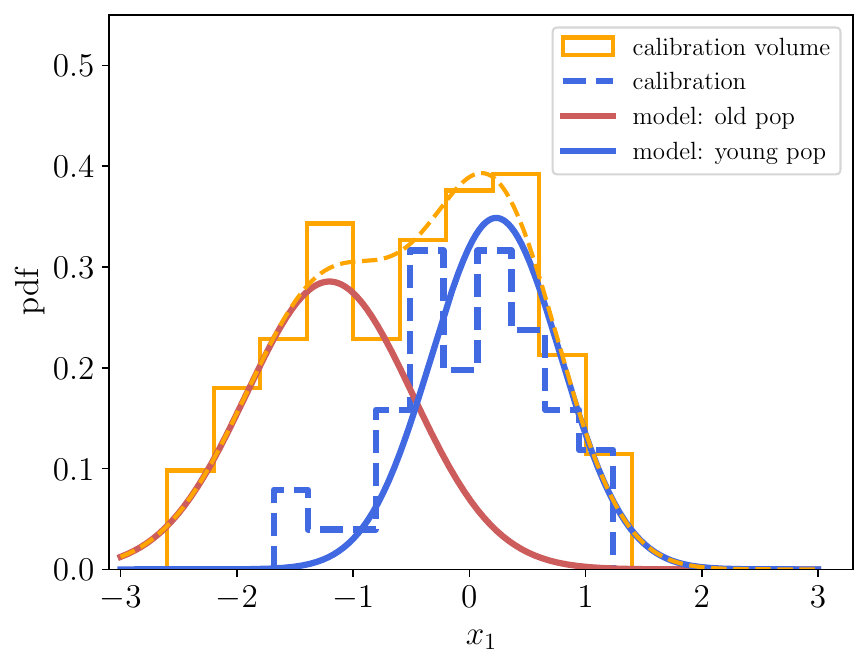} 
    \end{subfigure}
    \caption{Distribution of supernova stretch parameter in the Hubble flow \textit{(left)} and in the calibration galaxies \textit{(right)}. The stretch distribution in the Hubble flow exhibits 
    a bimodality which is well reproduced by a two-population Gaussian mixture model (the dashed line). The calibration supernovae are consistent with originating from the population associated with the high-stretch peak (young population). The right panel demonstrates this property by comparing the calibration sample to the stretch distribution for type Ia supernovae 
    found in the same comoving volume as the calibration galaxies, and to the corresponding two-population Gaussian mixture model. For the sake of better visual comparison, we keep 
    the same normalisation of the calibration sample and the young population from the calibration volume (right panel). The Hubble constant measurement presented in this work assumes that only the young supernova population (associated with the high-stretch peak of the stretch parameter distribution) is involved in propagating Cepheid distances to the Hubble flow.
    }
              \label{x1_hist}
    \end{figure*}

Figure~\ref{x1_hist} shows the key observational property motivating our two-population model. The data used in this study as well as a range of other type Ia supernova 
compilations, especially the volume limited sample from the Zwicky Transient Factory \citep{Ginolin2025}, reveal a clear bimodality in the distribution of stretch parameter 
$x_{1}$. This property can be effectively reproduced using a mixture of two supernova populations. The resulting populations may also differ in terms of the remaining intrinsic 
and extrinsic (dust and extinction related) supernova properties. This is expected based on a well studied observational relation between the supernova stretch and 
host-galaxy environment \citep{Sullivan2006,Rigault2013,Rigault2020,Nicolas2021}. Within this framework, 
the high-stretch supernova population originates primarily from young (star forming) 
and thus more dusty environments, whereas the old stretch analog is mostly associated with old and less dusty host galaxies. The associated stellar environments overlap 
in a way that low-star formation contribute to some extent to the high-stretch population but high-star formation analogs appear to be more strongly associated with the 
high-stretch population \citep{Rigault2020}. In our analysis, we assume that supernova local environments are differentiated based on population typing with 
respect to the stretch parameter. We will hereafter refer to the populations associated with the high- and low-stretch peak of the stretch distribution as young and old 
populations (with respectively "y" and "o" subscripts to denote them).

We model the two population as a sum of two prior distributions with distinct sets of hyperparameters, i.e.
\begin{equation}
p_{\rm prior}(\pmb{\phi})=f_{\rm o}p_{\rm prior,o}(\pmb{\phi}|\pmb{\Theta_{\rm o}})+(1-f_{\rm o})p_{\rm prior,y}(\pmb{\phi}|\pmb{\Theta_{\rm y}}),
\label{prob_main}
\end{equation}
where $f_{\rm o}\equiv 1-f_{\rm y}$ is the relative fraction of the old supernova population and it is a free parameter in modelling type Ia supernova 
in the Hubble flow. The primary difference between the old and young population lies in the mean stretch parameter and it is driven by the bimodality 
in the stretch distribution. As we shall see in the following section, further differences revealed in a full analysis includes other properties such as the mean 
absolute magnitude or the extinction parameter. The prior distribution which remains the same by construction in both populations is that related to distance moduli 
(given by cosmological model in the Hubble flow or the external measurements from Cepheid observations in the calibration galaxies).

Figure~\ref{x1_hist} compares the stretch parameter distribution in the calibration sample to that obtained from all supernova within the same local comoving volume 
(with maximum redshift of $0.0168$ corresponding to the most distant calibration galaxy). It is apparent that the distribution in the calibration sample coincides 
closely with the population associated with the high-stretch peak of the reference sample. In fact, we find that the calibration sample is statistically indistinguishable from 
the young population which is represented by a Guassian distribution with mean of $\mu=0.23$ and scatter $\sigma=0.56$, as determined from fitting a mixture of two Gaussians to the bimodal distribution of the reference sample. The two-sided KS test yields $p=0.024$, and $p=0.25$ when 4 (out of 42) calibration supernovae with $x_{1}<-0.8$ are omitted. 
This basic observational argument justifies the assumption that the calibration sample consists entirely of supernovae drawn from the young (associated with high-stretch peak) 
population. This is a direct implication of selecting late-type star-forming host galaxies for observations of Cepheids. We adopt this assumption, i.e. $f_{\rm o,cal}=0$, in our 
determination of the Hubble constant. All fits are tested for a possible impact of including or discarding the four supernovae with $x_{\rm 1}<-0.8$.

\section{Results}
\label{sec:results}

We obtain best-fit parameters by integrating the posterior probability using a \textit{Monte Carlo Markov Chain} method implemented in the \textit{emcee} code \citep{emcee}. 
Summary statistics of the best-fit models are provided in the form of posterior means and errors given by 16th and 84th percentiles of the marginalised probability distributions. 
In all figures, the $1\sigma$ and $2\sigma$ credible contours contain respectively 68 and 95 per cent of the corresponding two-dimensional marginalised probability 
distributions. The two-population model is primarily constrained by the supernova data from the Hubble flow. It is therefore instructive to begin with an overview of 
intrinsic and extrinsic properties of type Ia supernovae obtained from the Hubble flow data.

\subsection{Constraints from the Hubble flow}

 \begin{figure*}
   \centering
   \includegraphics[width=\linewidth]{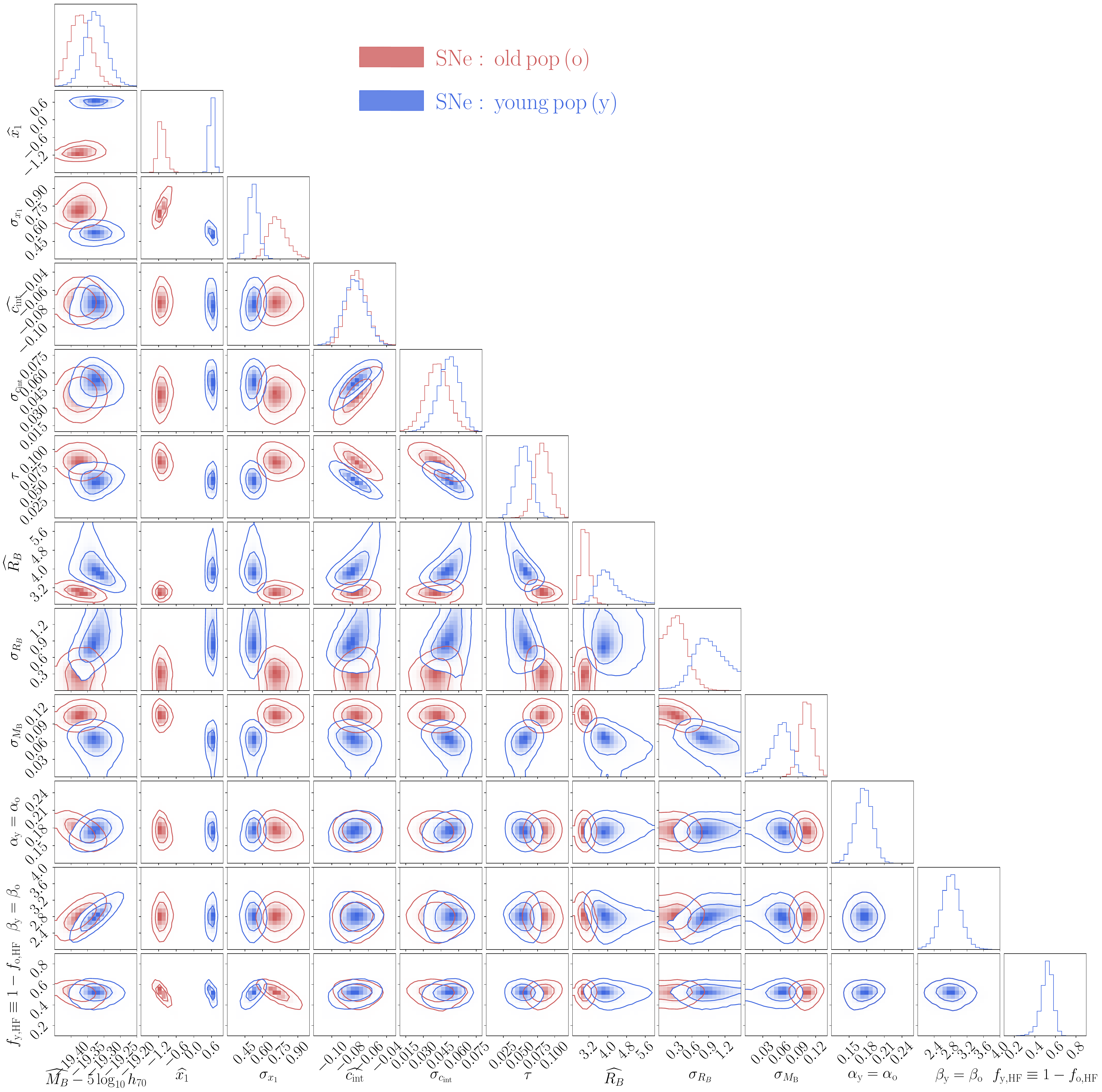}
    \caption{Constraints on hyperparameters of the two-population mixture model from the analysis of type Ia supernovae in the Hubble flow from the Pantheon+ compilation. The red 
    and blue colours denote respectively the old and young populations (associated with the low- and high-stretch peak of the stretch distribution, respectively). The contours show $1\sigma$ and $2\sigma$ credible regions containing 68 and 95 per cent 
    of two-dimensional marginalised probability distributions.}
              \label{baseline_HF}
    \end{figure*}

Figure~\ref{baseline_HF} shows constraints on the model parameters and Table~\ref{bestmodels_HF} provides the corresponding best-fit values and credible intervals. 
The two supernova populations contribute to the overall rate in the Hubble flow with comparable relative fractions, i.e. $f_{\rm o,HF}\approx f_{\rm y,HF}$. The main constraining 
power for differentiating between the supernova populations comes from the bimodality in the stretch parameter distribution. The obtained peaks and widths of the population 
distributions agree fairly well with those determined from fitting analogous stretch parameter distribution obtained from the volume limited sample from the Zwicky Transient Factory \citep{Ginolin2025}, although a $1.7\sigma$ difference is found for $\widehat{X_{1}}$ of the young population (with $\widehat{X_{1}}=0.42\pm0.08$ from the ZTF).

\begin{figure*}
    \centering
    \begin{subfigure}[l]{0.43\textwidth}
        \centering
        \includegraphics[width=\linewidth]{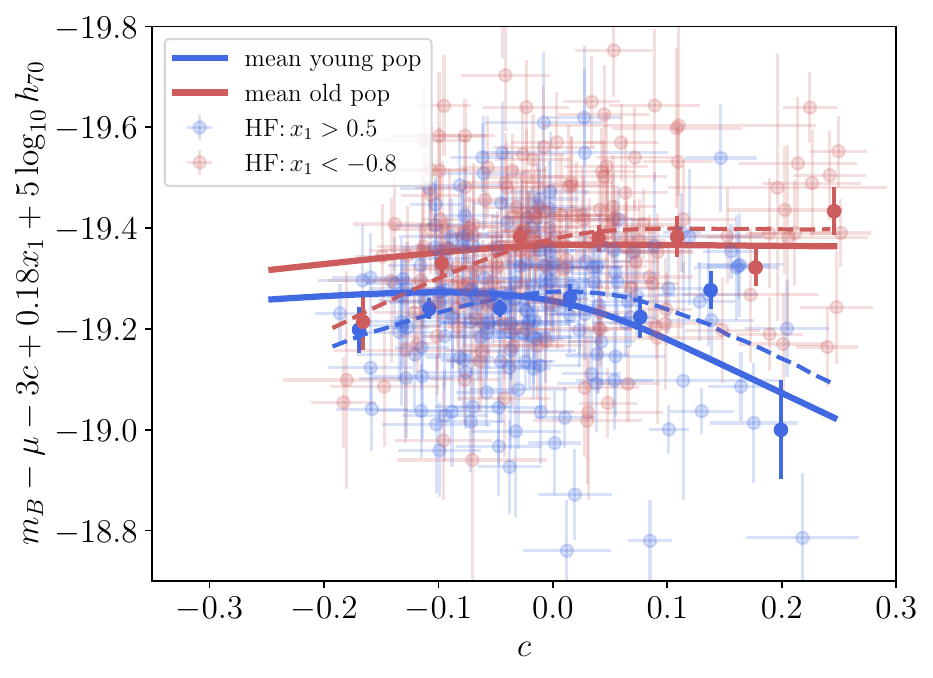} 
    \end{subfigure}
    \begin{subfigure}[r]{0.47\textwidth}
        \centering
        \includegraphics[width=\linewidth]{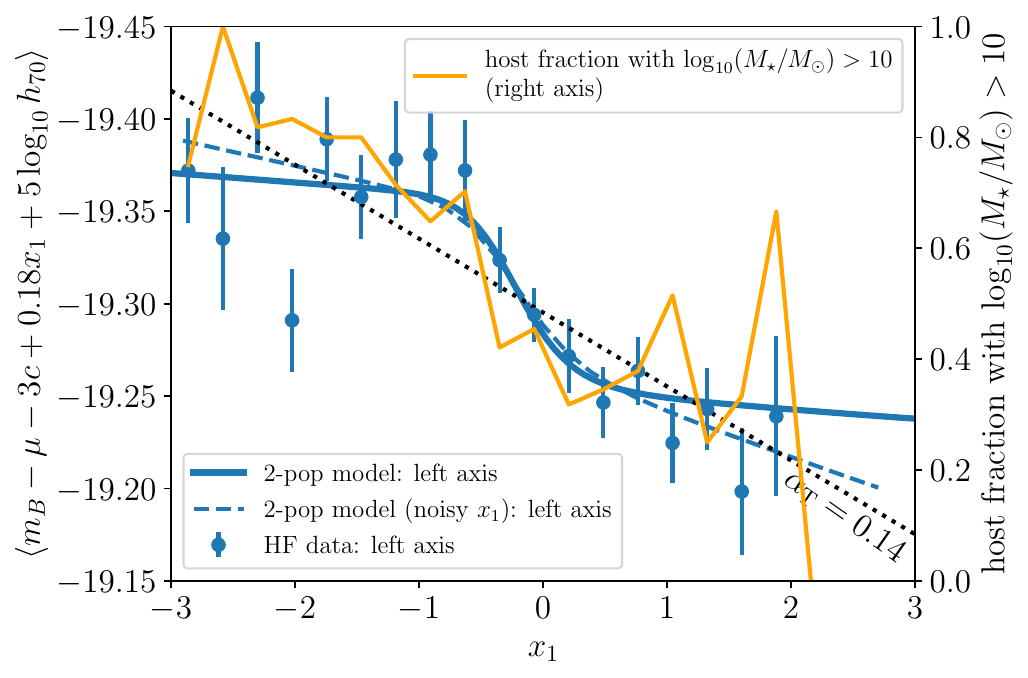} 
    \end{subfigure}
    \caption{Comparison between the best-fit two-population model and the supernova data in the Hubble flow. The panels show supernova absolute magnitudes obtained 
    from the standard linear corrections in stretch and supernova colour, as a function of the colour parameter \textit{(left)} and the stretch parameter \textit{(right)}. For a more 
    direct comparison with the noisy data, the dashed lines show the best-fit model predictions which take into account noise due to measurement uncertainties (omitted for the solid lines). The data on the left panel are restricted to supernovae with $x_{1}>0.5$ or $x_{1}<-0.8$ representing high-purity samples of the underlying young and old populations. The apparent difference between the slopes of the colour correction at $c>0$ in the two populations reflects the difference in the derived mean extinction coefficients $R_{\rm B}$. The right y-axis of the right panel shows the fraction of supernova host galaxies with $\log_{10}(M_{\star}/M_{\odot})>10$. This demonstrates a close connection (although not equivalence) between the two-population model and the traditional mass-step correction \citep[see also][]{Briday2022, Wiseman2023}. The old population is dominated by high stellar-mass hosts, whereas the young population is widely distributed over host-galaxy stellar masses.
    }
\label{data_comparison}
\end{figure*}

The two supernova populations differ also in terms of extinction and intrinsic luminosity. Supernovae from the young population are intrinsically fainter ($\Delta \widehat{M_{\rm B}}\approx 0.04$) and subject to a stronger total-to-selective extinction ($\Delta \widehat{R_{\rm B}}\approx 1$). The $(\widehat{M_{\rm B}},\widehat{R_{\rm B}})$ panel in Figure~\ref{baseline_HF} shows that the population difference in terms of these two parameters reaches $4\sigma$ statistical significance \citep[comparable to or stronger 
than significance of measuring the mass-step correction in unrestricted cosmological supernova samples, see e.g.][]{Scolnic2018}. The derived mean extinction coefficient 
for the young population is consistent with a typical value of $R_{\rm B}=4.3$ measured in the Milky Way \citep{Fitzpatrick2007,Schlafly2016} and thus implies similar dust properties 
as those in the Milky Way.  On the other hand, $\widehat{R_{\rm B}}\approx 3$ obtained for the old population is clearly discrepant with the Milky Way-like dust composition. The extinction coefficients obtained for the two supernova populations agree well with the estimates found for analogous young and old populations differentiated by the supernova progenitor age derived from forward modelling of host-galaxy properties \citep{Wiseman2022}. 

Keeping in mind that the young (old) supernova populations dominate at low (high) stellar masses of host galaxies, we find a close correspondence between our results and independent constraints 
from \citet{Grayling2024} based on the hierarchical Bayesian spectral energy distribution (SED) model BayeSN \citep{Mandel2022,Thorp2021}. The study found that type Ia supernovae in high-mass host galaxies are intrinsically brighter ($\Delta M_{\rm B}\approx -0.04$~mag) with a weaker extinction ($R_{\rm B}\approx 3.3-3.5$) than their analogs in low-mass host galaxies ($R_{\rm B}\approx 3.7-4.4$). We notice also a close match between our estimates of $R_{\rm B}$ and those obtained by \citet{Popovic2021} when comparing our young (old) supernova populations to those in low (high)-mass host galaxies.

The obtained reddening scale for the old population ($\tau\equiv\langle E(B-V)\rangle\approx 0.09$) exceeds the expected interstellar reddening in early type host galaxies which dominate 
in this supernova population \citep{Hallgren2025}. This discrepancy raises a conjecture that the effective extinction model for the old population is caused by interstellar dust. Although $R_{\rm B}\approx 3$ and its unknown physical origin were encountered before in a range of studies \citep{Brout2021,Popovic2021,Burns2014}, our results suggest that this problem is restricted primarily to the old supernova population. The negligible difference between $\widehat{R_{\rm B}}$ and $\beta$ found in this population suggests that the magnitude-colour relation may be driven by the same physical mechanism across the entire range of colours. This would mean that the exponential component of the observed colour distribution arises likely from a larger and asymmetric variation in intrinsic colours than the assumed Gaussian prior. The picture is consistent with the fact that intrinsically red type Ia supernovae in early-type galaxies or in the regime of low values of the stretch parameter are commonly observed \citep{Burns2014,Gonzalez2011}.

\begin{table*}
\caption{Best-fit hyperparameters of the prior probability distributions of the two-population mixture model.}
\begin{center}
\begin{tabular}{lcccc}
\hline
 & \multicolumn{2}{c}{Hubble flow} & \multicolumn{2}{c}{Hubble flow + calibration sample} \\
 & old pop (low $X_{1}$) & young pop (high $X_{1}$) & old pop (low $X_{1}$) & young pop (high $X_{1}$) \\
\hline
 \\
$\widehat{M}_{\rm B}$ & $ -19.37 ^{+ 0.03 }_{- 0.03 }\,(h=0.7)$ & $ -19.33 ^{+ 0.03 }_{- 0.03 }\,(h=0.7)$ & $ -19.34 ^{+ 0.05 }_{- 0.05 }$ & $ -19.28 ^{+ 0.04 }_{- 0.04 }$ \\
$\widehat{X}_{1,{\rm HF}}$ & $ -1.09 ^{+ 0.13 }_{- 0.13 }$ & $ 0.60 ^{+ 0.07 }_{- 0.07 }$ & $ -1.12 ^{+ 0.11 }_{- 0.11 }$ & $ 0.59 ^{+ 0.06 }_{- 0.06 }$ \\
$\sigma_{X_{1},{\rm HF}}$ & $ 0.73 ^{+ 0.08 }_{- 0.08 }$ & $ 0.52 ^{+ 0.04 }_{- 0.04 }$ & $ 0.72 ^{+ 0.07 }_{- 0.07 }$ & $ 0.53 ^{+ 0.04 }_{- 0.04 }$ \\
$\widehat{c_{\rm int}}$ & $ -0.072 ^{+ 0.011 }_{- 0.011 }$ & $ -0.075 ^{+ 0.012 }_{- 0.012 }$ &  $ -0.074 ^{+ 0.011 }_{- 0.011 }$ & $ -0.067 ^{+ 0.012 }_{- 0.012 }$ \\
$\sigma_{c_{\rm int}}$ & $ 0.041 ^{+ 0.010 }_{- 0.010 }$ & $ 0.051 ^{+ 0.009 }_{- 0.009 }$ & $ 0.039 ^{+ 0.010 }_{- 0.010 }$ & $ 0.059 ^{+ 0.008 }_{- 0.008 }$ \\
$\tau$ & $ 0.082 ^{+ 0.011 }_{- 0.011 }$ & $ 0.055 ^{+ 0.011 }_{- 0.012 }$ & $ 0.083 ^{+ 0.011 }_{- 0.011 }$ & $ 0.050 ^{+ 0.011 }_{- 0.012 }$\\
$\widehat{R_{\rm B}}$ & $ 3.030 ^{+ 0.177 }_{- 0.177 }$ & $ 4.074 ^{+ 0.510 }_{- 0.494 }$ & $ 3.018 ^{+ 0.169 }_{- 0.171 }$ & $ 4.281 ^{+ 0.676 }_{- 0.610 }$ \\
$\sigma_{R_{\rm B}}$ & $ 0.323 ^{+ 0.191 }_{- 0.207 }$ & $ 0.915 ^{+ 0.313 }_{- 0.279 }$ &  $ 0.316 ^{+ 0.188 }_{- 0.202 }$ & $ 0.934 ^{+ 0.329 }_{- 0.297 }$ \\
$\sigma_{M_{\rm B}}$ & $ 0.104 ^{+ 0.011 }_{- 0.011 }$ & $ 0.059 ^{+ 0.017 }_{- 0.017 }$ & $ 0.104 ^{+ 0.011 }_{- 0.011 }$ & $ 0.051 ^{+ 0.018 }_{- 0.018 }$ \\
$\alpha$ & \multicolumn{2}{c}{$ 0.175 ^{+ 0.014 }_{- 0.014 }$} & \multicolumn{2}{c}{$ 0.185 ^{+ 0.012 }_{- 0.012 }$} \\
$\beta$ & \multicolumn{2}{c}{$ 2.797 ^{+ 0.229 }_{- 0.231 }$} & \multicolumn{2}{c}{$ 2.821 ^{+ 0.181 }_{- 0.182 }$} \\
$f_{\rm o,HF}=1-f_{\rm y,HF}$ & \multicolumn{2}{c}{$ 0.484 ^{+ 0.054 }_{- 0.055 }$} & \multicolumn{2}{c}{$ 0.471 ^{+ 0.047 }_{- 0.047 }$} \\
$f_{\rm o,cal}=1-f_{\rm y,cal}$ & \multicolumn{2}{c}{-} & \multicolumn{2}{c}{$\equiv 0$} \\
\end{tabular}
\label{bestmodels_HF}
\end{center}
\tablefoot{The results were obtained from fitting type 
Ia supernovae in the Hubble flow (left) or from a joint analysis of the Hubble flow and calibration samples (right). Best-fit results are provided as the posterior mean values and errors given by credible intervals containing 68 per cent of the marginalised probabilities. Two parameters ($\alpha$ and $\beta$) are assumed to be the same in both supernova populations. 'HF' and 'cal' denote parameters used exclusively in the Hubble flow or the calibration sample.}
\end{table*}

Fitting $\alpha$ and $\beta$ coefficients independently in the two populations yields mutually consistent constraints on these parameters. For this reason, all models presented 
in this work assume that these coefficients are the same in both supernova populations. The best-fit $\alpha$ (0.175) is noticeably higher than a typical value derived from 
fitting the Tripp calibration formula \citep[$\alpha_{\rm T}\approx0.14$, see e.g. ][]{Bet2014}. This difference is directly related to the magnitude 
offset between the two populations which is captured by our model, but not accounted for by the Tripp correction. For supernova samples dominated by one of the two populations, we expect $\alpha_{\rm T}$ measured from fitting the Tripp calibration formula to match our estimate. This may be the case of high-redshift supernova samples where one can expect a higher young-to-old population ratio \citep[see the evolution argument from][]{Nicolas2021}. Interestingly, the best-fit Tripp calibration obtained for high-redshift supernovae from the Dark Energy Survey (DES) results in $\alpha_{\rm T}=0.17$ \citep{Vincenzi2024}.

The left panel of Figure~\ref{data_comparison} compares the mean Tripp-like corrected magnitude as a function of the colour parameter to the best-fit two-population model. 
The data are selected from the high- and low-stretch tails of the $x_{1}$ distribution so that the resulting two supernova samples represent accurately the two populations. 
The apparent separation of the mean magnitude-colour relations for the two populations is effectively reproduced by different $\widehat{M_{\rm B}}$ and $\widehat{R_{\rm B}}$ 
measured in both supernova populations. Different total-to-selective extinction coefficients in both populations manifest themselves as divergent slopes at colours driven 
by reddening ($c\gtrsim 0$). The apparent steepening of the slope in the data at blue colours ($c\lesssim -0.1$) is an effect of noise from non-negligible measurement 
errors \citep[see e.g. Appendix A of][]{Ginolin2025}. It is well reproduced by a Monte Carlo sampling of our best-fit model which includes a Gaussian noise in the observed 
colour $c$ based on the mean measurement error from the data (see the dashed lines). We note that our likelihood incorporates the measurement errors (and correlations) 
of light curve parameters of each supernova so that best-fit models show the true physical properties. The measurement noise is simulated only for the purpose of 
comparison between the best-fit models and the noisy data.

The net effect of differences between the two supernova populations in terms of $\widehat{M_{\rm B}}$ and $\widehat{R_{\rm B}}$ can be effectively thought of as a step correction 
in the stretch parameter. This is shown on the right panel of Figure~\ref{data_comparison} which compares the mean Tripp-based corrected magnitude as a function of $x_{1}$ to 
the corresponding mean magnitude computed for our best-fit two-population model. We can see that a continuous transition between high- and low-stretch supernovae, which 
in the traditional modelling would requires adjusting a priori the unknown sharpness of the step correction, emerges naturally from a partial overlap of the two populations. 
An attempt to model the transition employing a single population model without any mass step in $x_{1}$ would necessitate using a non-linear magnitude-stretch relation 
\citep{Ginolin2024a}. The traditional Tripp calibration formula with $\alpha_{\rm T}=0.14$ provides a linear approximation of the observed magnitude-stretch relation (see the black dashed line on the right panels of Figure~\ref{data_comparison}).

As shown on the right panel of Figure~\ref{data_comparison}, the mean Tripp-like corrected magnitude as a function of $x_{1}$ coincides closely with the fraction of massive ($M_{\star}>10^{10}M_{\odot}$) host galaxies as a function $x_{1}$. This correlation is not surprising and it reflects mutual relations between supernova stretch parameter, 
host-galaxy stellar environment and stellar mass. The stellar mass of $10^{10}M_{\odot}$ employed traditionally as a transition mass of the mass step correction is actually 
the mass scale which maximises the difference between relative fractions of old and young stellar environments \citep[see e.g.][]{Kauffmann2003}. 
From this point of view, the mass step correction can be thought of as an emergent relation which captures only partially the actual differences between environment-dependent 
properties of type Ia supernovae. Although it is instructive to think about the correspondence between supernova populations defined by the stretch parameter (as a proxy of 
supernova host-galaxy stellar population age) and the host stellar mass \citep{Brout2021}, these two ways of population modelling are far from being equivalent and 
result in different magnitude steps between supernova populations \citep{Briday2022}. While the old supernova population is dominated by high stellar-mass host galaxies 
(80 per cent), the young population includes both low and high stellar-mass host galaxies with comparable relative fractions. These properties can be modelled using 
supernova progenitor age as the primary variable \citep{Wiseman2023}.

The two supernova populations exhibit different properties of derived scatter, as encapsulated in $\sigma_{M_{\rm B}}$ and $\sigma_{\rm R_{\rm B}}$ (see Table~\ref{bestmodels_HF}). 
The old population is driven primarily by achromatic scatter with $\sigma_{\rm int}\approx 0.1$, whereas the scatter in the young population increases with supernova colour. The chromatic 
component of the scatter in the young supernova population is ascribed in the assumed model to scatter in $R_{\rm B}$. The best-fit $\sigma_{\rm R_{\rm B}}$ found in 
the young population is about 3 times larger than what is measured in the Milky Way \citep{Fitzpatrick2007}. This may signify either a substantially larger range of dust 
properties than those in the Milky Way or an artefact of incomplete separation of supernova populations. Since the stretch parameter is only a proxy of the actual stellar 
environment \citep[as measured by the specific star formation rate;][]{Rigault2020}, the latter scenario is quite plausible. In order to assess the impact of the residual 
presence of the old supernova population in the high-stretch component of the stretch distribution on model parameter estimation, we modify our model in a way that the old 
population is allowed to contribute partially to the high-stretch component and its relative fraction is an additional free parameter. The results are shown in Appendix B. We find 
that about 25 per cent (maximum marginalised posterior, see Figure~\ref{baseline_HF_ref}) of the high-stretch component comes from the old supernova population. This 
estimate agrees with the fraction of high-stretch supernovae observed in elliptical galaxies, e.g. 24 per cent for $x_{1}>1.0$ and 26 per cent for $x_{1}>0.5$ estimated from the ZTF data \citep{Senzel2025}. The best-fit parameters of both populations remain nearly unchanged. We notice, however, a slight increase of the mean extinction parameter 
in the young supernova population to $\widehat{R_{\rm B}}=4.6$. Constraints on $\sigma_{\rm R_{\rm B}}$ and $\sigma_{\rm M_{\rm B}}$ move towards smaller values indicating more homogeneous properties of the refined young supernova population.

\subsection{Hubble constant}

We measure the Hubble constant from a joint likelihood combining the Hubble flow (explored in the previous subsection) and the 
calibration data blocks. All fits involve modelling both supernova populations in 
the Hubble flow, although only the young population has an effect on the Hubble constant. Distance moduli of 37 calibration galaxies are free nuisance parameters with the prior probability given 
by SH0ES measurements obtained from the Cepheid observations. Based on the arguments put forward in section~\ref{sect:twopop}, 
we assume that the calibration sample consists exclusively of supernovae of the young population ($f_{\rm o,cal}\equiv 0$). The mean and dispersion of $x_{1}$ in the calibration sample are fitted independently of those in the Hubble flow. This accounts for a small but apparent difference between the distribution of $x_{1}$ in the young population from the local comoving volume including the calibration galaxies ($\mu_{x_{1}}\approx 0.2$ and $\sigma_{\rm x_{1}}\approx 0.6$) and from 
the Hubble flow ($\mu_{x_{1}}\approx 0.6$ and $\sigma_{\rm x_{1}}\approx 0.5$). A part of the apparent shift in the stretch distribution may be ascribed 
to a difference between host galaxy stellar masses, with higher $x_{1}$ expected for less massive host galaxies \citep{Ginolin2024a}. 
Indeed, we find that the mean host galaxy stellar mass for the young supernova population in the Hubble flow ($x_{1}>0.5$) is about $0.5$~dex 
smaller than in the calibration sample (based on stellar mass estimates from the Pantheon+ catalogue).

We begin our analysis with the most conservative assumption that all parameters of the young supernova population 
(except $\mu_{\rm x_{1}}$ and $\sigma_{\rm x_{1}}$) are the same in the Hubble flow and the calibration sample (model A). 
The results are shown in Tables~\ref{bestmodels_HF}-\ref{bestH0}. The best-fit Hubble constant is $H_{0}=71.45\pm1.03$~km~s$^{-1}$~Mpc$^{-1}$. 
The uncertainty matches very well that obtained by \citet{Riess2022}, but the the best-fit value is $1.5\sigma$ lower than the SH0ES result. 
Constraints on the parameters of the two-population model remain virtually unchanged relative to those based on the Hubble 
flow data (see Table~\ref{bestmodels_HF}). This demonstrates that the main constraining power comes from supernovae in the Hubble flow.

One cannot rule out a priori that the calibration sample may differ to some degree from the young population in the Hubble flow in terms of 
extrinsic properties or effective scatter. Possible differences may result from unequal distributions of dust column density (and thus reddening) 
or imprecise match of supernova environments. Fits with independent $\sigma_{M_{\rm B}}$ and $\sigma_{\rm R_{\rm B}}$ in the calibration 
sample yield merely upper bounds on these two parameters (with $1\sigma$ upper limits of $\sigma_{M_{\rm B}}<0.066$ and $\sigma_{R_{\rm B}}<0.93$) 
suggesting that the calibration supernovae are consistent with 
vanishing dispersion both in the absolute magnitude $M_{\rm B}$ and $R_{\rm B}$. It is worth noting that $\sigma_{M_{\rm B}}=0$ and 
$\sigma_{R_{\rm B}}=0$ do not imply vanishing scatter around the mean, Tripp calibration-based corrected peak magnitude as a function of the colour parameter. 
One can show that a scatter of $0.05$~mag at $c=0$ arises purely from a difference between best-fit $\beta$ and $\widehat{R_{\rm B}}$. 
We find that fits with $\widehat{R_{\rm B}}$ measured independently in the calibration sample yields values which are fully consistent with the 
standard Milky-Way extinction, irrespective of the assumption about $\sigma_{M_{\rm B}}$ and $\sigma_{R_{\rm B}}$. We find 
$\widehat{R_{\rm B}}=3.99\pm0.90$ for $\sigma_{M_{\rm B}}$ and $\sigma_{\rm R_{\rm B}}$ let to be independent free parameters 
in the calibration sample, and $\widehat{R_{\rm B}}=4.88\pm0.52$ for $\sigma_{M_{\rm B}}\equiv 0$ and $\sigma_{R_{\rm B}}\equiv 0$ in the calibration sample.

\begin{figure}
        \centering
        \includegraphics[width=\linewidth]{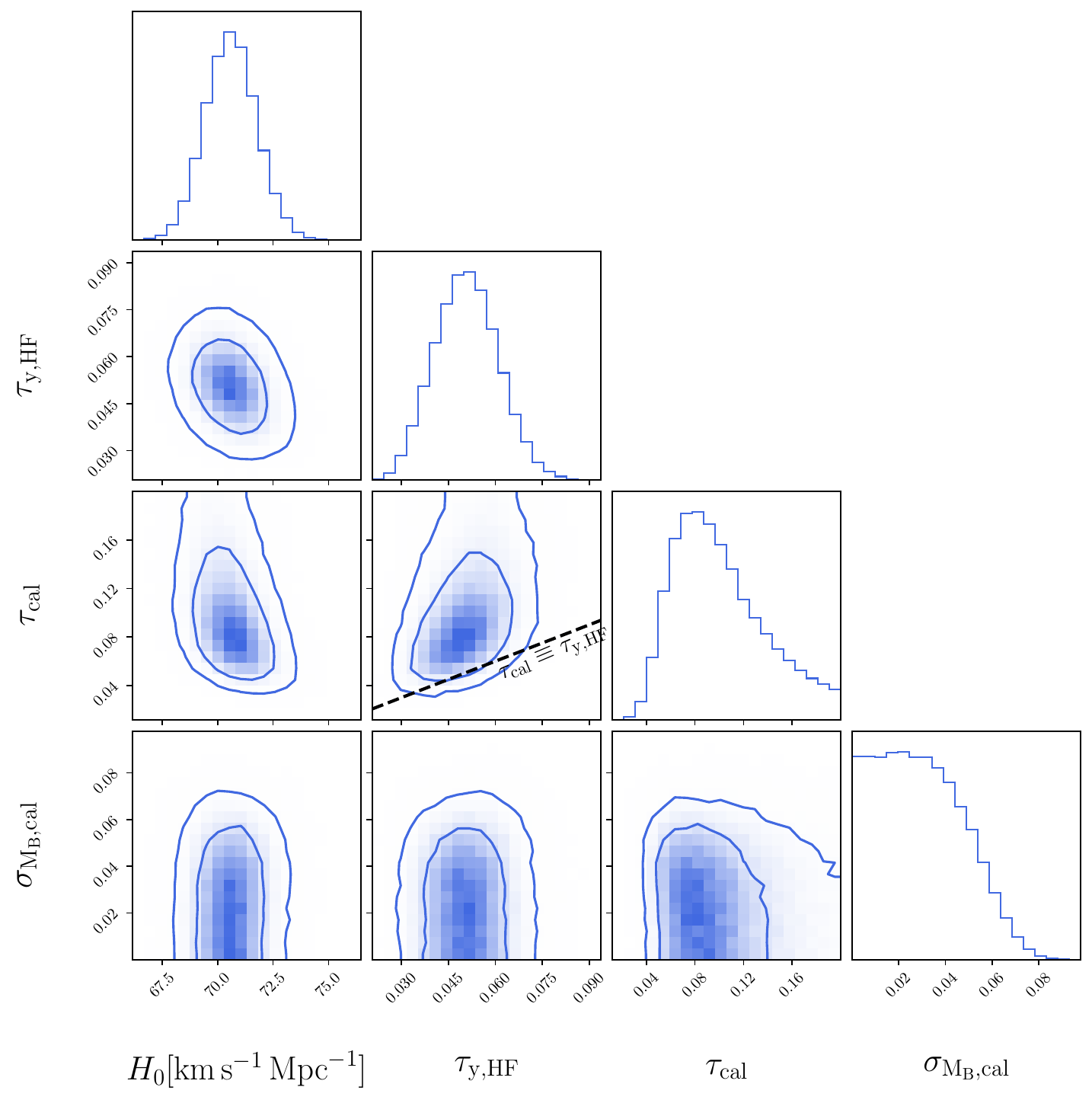} 
    \caption{Constraints on the Hubble constant and selected parameters in model B assuming independent scales of host-galaxy
    interstellar reddening $\tau$ and scatter $\sigma_{M_{\rm B}}$ in the young supernova population of the Hubble flow (HF) and the calibration (cal) galaxies, and a fixed, Milky-Way like distribution of $R_{\rm B}$ in the calibration galaxies. The best-fit model demonstrates a reduction of the Hubble constant due to a larger reddening scale in the 
    calibration sample than in the Hubble flow ($\tau_{\rm cal}>\tau_{\rm y,HF}$). The contours show $1\sigma$ and $2\sigma$ 
    credible regions containing 68 and 95 per cent of 2-dimensional marginalised probability distributions.
    }
\label{corner_model_B}
\end{figure}

As a way to explore the impact of constraining the reddening distribution in the calibration supernovae directly from the calibration 
data, we consider a model in which $\tau$ of the young population is independent in the calibration sample and the 
Hubble flow (model B). Despite the fact that the calibration sample is consistent with $\sigma_{M_{\rm B}}=0$, we take a less restrictive approach and let the 
parameter vary so that its values is always smaller than $\sigma_{M_{\rm B}}$ in the young population of the Hubble flow. We also assume a fixed distribution 
of $R_{\rm B}$ with $\widehat{R_{\rm B}}=4.3$ and $\sigma_{\rm R_{\rm B}}=0.4$ based on observational constraints from the Milky Way \citep{Fitzpatrick2007,Schlafly2016,Legnardi2023}. The assumed Milky Way-like distribution is consistent with $\sigma_{\rm R_{\rm B}}$ measured in the calibration sample and it is also theoretically motivated as the most plausible model for interstellar extinction in galaxies which are close analogs of the Milky Way.

As shown in Figure~\ref{corner_model_B}, fitting the new model (model B) to the data yields a lower value of the Hubble 
constant as a result of an elevated reddening scale $\tau$ derived from the calibration data. Table~\ref{bestH0} summarises our results. 
For a more physical interpretation of $\tau$ in the calibration sample, the table provides also derived mean reddening conditioned 
on the range of observed supernova colours (the actual mean reddening in the calibration sample). The resulting Hubble constant 
is $H_{0}=70.58\pm1.15$~km~s$^{-1}$~Mpc$^{-1}$. Its best-fit value is about $1\sigma$ smaller than in model A and 
the modelled uncertainty is 10 per cent larger. Allowing for an unrestricted range of $\sigma_{M_{\rm B}}$ in the calibration sample 
results in slightly larger error but virtually the same best-fit value: $H_{\rm 0}=70.66\pm1.19$~km~s$^{-1}$~Mpc$^{-1}$. We also checked that model B conditioned 
on $\tau_{\rm cal}\equiv\tau_{\rm HF}$ yields $H_{0}=71.2\pm1.1$~km~s$^{-1}$~Mpc$^{-1}$ 
and thus recovers closely the measurement from model A. This means that the lower estimate of the Hubble constant in model B is primarily driven by a higher reddening scale derived from the calibration sample.

\begin{figure}
        \centering
        \includegraphics[width=\linewidth]{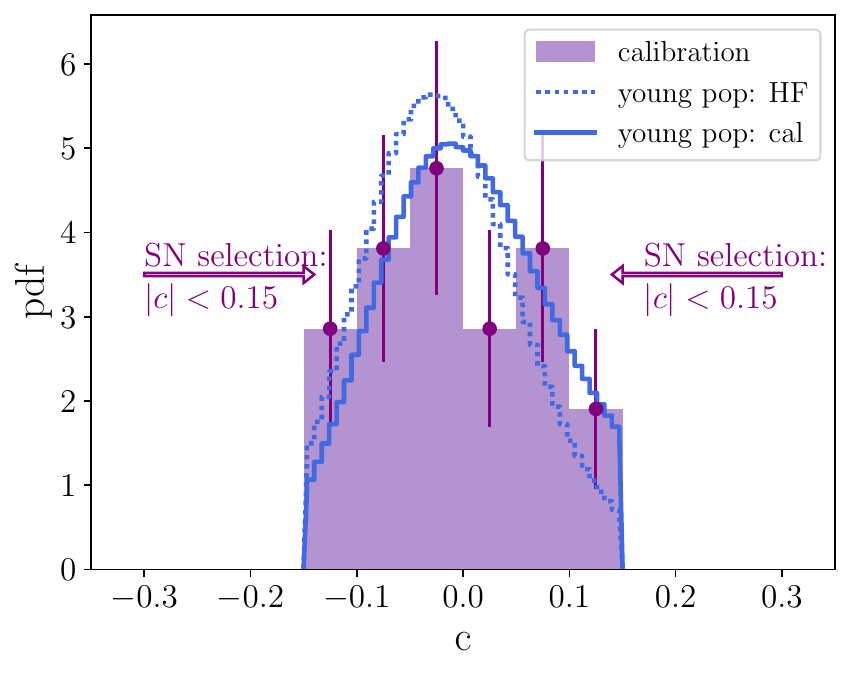} 
    \caption{Comparison between the distribution of supernova colour parameters in the calibration sample and the best-fit model with 
    the reddening scale $\tau$ measured either from the Hubble flow supernovae (model A) or from the calibration supernovae (model B). 
    The best-fit reddening scale in the calibration sample is larger than in the Hubble flow and this preference is driven by 
    a higher fraction of reddened supernovae ($c\sim 0.1$) than in the corresponding young population from the Hubble flow.
    }
\label{c-calibration}
\end{figure}

\begin{table*}
\caption{Best-fit Hubble constant and extinction-related parameters of the young supernova population.}
\centering
\begin{tabular}{lccccccccccc}
\hline
model & $H_{0}$ & \multicolumn{2}{c}{$\widehat{R_{\rm B}}$} & \multicolumn{2}{c}{$\sigma_{R_{\rm B}}$} & \multicolumn{2}{c}{$\sigma_{M_{\rm B}}$} & \multicolumn{2}{c}{$\tau$} & $\langle E(B-V)\rangle$ \\ 
 &  & \multicolumn{2}{c}{} & \multicolumn{2}{c}{} & \multicolumn{2}{c}{} & \multicolumn{2}{c}{} & $|c|<0.15$ \\
\hline
 & & HF-pop y & cal & HF-pop y & cal & HF-pop y & cal & HF-pop y & cal & cal\\
\hline
 \\
model A & $71.45^{+1.03}_{-1.03}$ & \multicolumn{2}{c}{$4.28^{+0.68}_{-0.61}$}  &  \multicolumn{2}{c}{$0.93^{+ 0.33 }_{- 0.30 }$} &  \multicolumn{2}{c}{$ 0.051^{+ 0.018 }_{- 0.018 }$} &  \multicolumn{2}{c}{$ 0.050 ^{+ 0.011 }_{- 0.012 }$} & $0.0466^{+0.009}_{-0.010}$\\
 \\
model B & $70.59^{+1.15}_{-1.16}$ & $4.22^{+0.53}_{-0.51}$ & $\equiv 4.3$ & $0.91^{+0.31}_{-0.28}$ & $\equiv 0.4$ & $0.060^{+0.015}_{-0.015}$ & $<0.038$ & $0.051^{+0.010}_{-0.010}$ & $0.101^{+0.040}_{-0.037}$ & $0.071^{+0.015}_{-0.015}$ \\
\\
\hline
\multicolumn{11}{c}{omitting 4 SNe with $x_{1}<-0.8$ in the calibration sample} \\
\hline
\\
model A  & $71.20^{+1.05}_{-1.05}$ & \multicolumn{2}{c}{$4.37^{+0.69}_{-0.62}$}  &  \multicolumn{2}{c}{$0.94^{+ 0.34 }_{- 0.31 }$} &  \multicolumn{2}{c}{$ 0.050^{+ 0.018 }_{- 0.019 }$} &  \multicolumn{2}{c}{$ 0.049 ^{+ 0.011 }_{- 0.011 }$} & $0.0456^{+0.009}_{-0.009}$\\
 \\
model B  & $70.19^{+1.15}_{-1.15}$ & $4.22^{+0.55}_{-0.52}$ & $\equiv 4.3$ & $0.91^{+0.32}_{-0.29}$ & $\equiv 0.4$ & $0.060^{+0.015}_{-0.015}$ & $<0.036$ & $0.051^{+0.010}_{-0.010}$ & $0.120^{+0.045}_{-0.041}$ & $0.077^{+0.013}_{-0.013}$ \\
 \\
 \hline
\end{tabular}
\label{bestH0}
\tablefoot{The extinction properties are either assumed to be the same along the cosmic distance ladder (model A) or allowed to differ (model B) between the Hubble flow (HF) and the calibration sample (cal). 
The results are summarised as the posterior mean values and errors given by credible intervals (or upper limit in the case of $\sigma_{M_{\rm B}}$ in the calibration sample) containing 68 per cent of the marginalised probabilities. Two parameters 
in model B are fixed and their values are provided with the sign '$\equiv$'. The last column shows the mean reddening in the calibration sample conditioned on the selected 
range of colours $|c|<0.15$.}
\end{table*}

Figure~\ref{c-calibration} compares the best-fit colour distribution in the calibration sample obtained for models A and B to the data. 
In order to account for the effect of measurement uncertainties (which is incorporated in the likelihood but not in the best-fit distributions), 
the best-fit models shown in the figure are convolved with 
a Gaussian distribution with dispersion given by the mean error of the colour parameters in the calibration sample ($\sigma=0.024$). The figure shows that 
the preference for higher $\tau$ in the calibration sample is driven by a higher fraction of reddened supernovae ($c\sim 0.1$) than in the young population from the Hubble flow (represented by model A).

The apparent difference between the reddening scale derived from the Hubble flow and the calibration sample implies a non-vanishing offset between the mean reddening 
in the Hubble flow and the calibration sample. The offset cannot be eliminated by applying the same colour cuts in the Hubble flow, but it can be 
mitigated by selecting minimally reddened supernovae \citep[see also][]{Gall2024}. We find that the difference between the mean reddening of the young population in the calibration 
sample and the Hubble flow is $\Delta \langle E(B-V)\rangle=0.024$ for $|c|<0.15$ and $\Delta \langle E(B-V)\rangle=0.015$ for $c=0$. The difference between the best-fit 
Hubble constant in model A and B ($\delta m_{\rm B}=\delta \mu=5\log_{10}(H_{\rm 0,\,model\,A}/H_{\rm 0,\,model\,B})\approx0.027$) is primarily driven by the corresponding 
change of the mean reddening and intrinsic colour at $c=0$. With $\Delta \langle E(B-V)\rangle|(c=0)=0.015$ and $\Delta \langle c_{\rm int}\rangle|(c=0)=-0.012$, 
where '$|$' indicates conditioning on supernova colour $c$, we find $\delta m_{\rm B}=\beta \Delta \langle c_{\rm int}\rangle+\widehat{R_{\rm B}}\Delta \langle E(B-V)\rangle\approx 0.025$

As shown in section~\ref{sect:twopop}, 4 calibration supernovae with the lowest stretch parameter ($x_{1}<-0.8$) are not fully consistent 
with the extent of stretch parameter of the young supernova population (assuming symmetric Gaussian distribution). We find that 
omitting these supernovae from the fits results in a decrease of the best-fit Hubble constant by $0.3$~km~s$^{-1}$~Mpc$^{-1}$ (see Table~\ref{bestH0}). 
The result for model B (see above) is $H_{\rm 0}=70.19\pm1.15$~km~s$^{-1}$~Mpc$^{-1}$ and it is only weakly ($2.2\sigma$) discrepant with 
the Hubble constant measured from Planck data assuming a flat $\Lambda$CDM cosmological model. We checked that employing the model in which the old population in the Hubble flow is allowed 
to contribute to the high-stretch peak of the stretch distribution (see Appendix \ref{sec:appB}) results in a slight decrease of the Hubble constant estimate by $0.4$~km~s$^{-1}$~Mpc$^{-1}$ 
for model A and virtually no change of the measurement for model B.

\section{Discussion and summary} 
\label{sec:summary}

\begin{figure}
        \centering
        \includegraphics[width=\linewidth]{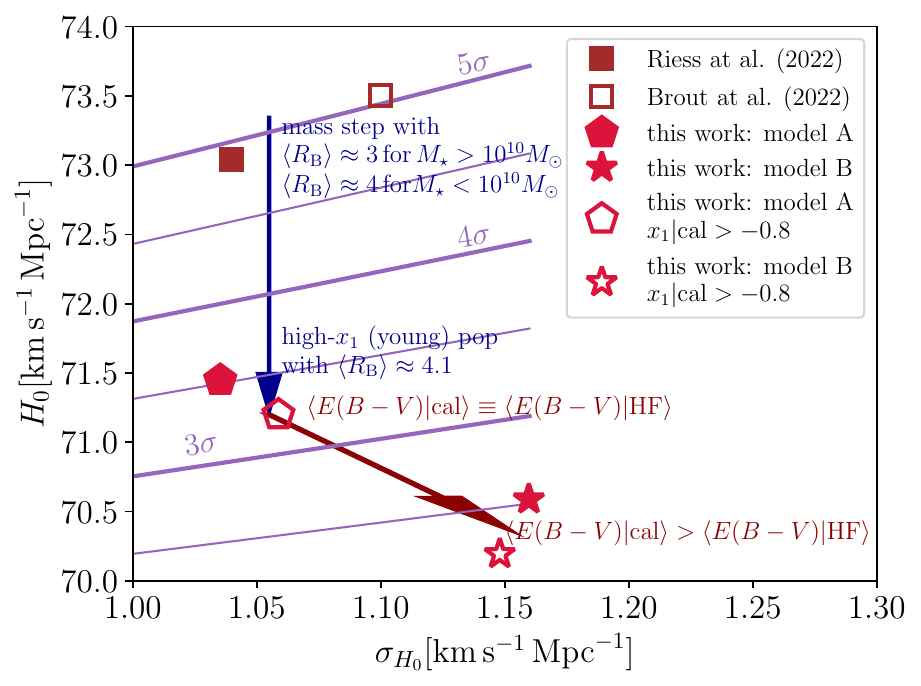} 
    \caption{Best-fit Hubble constant and its uncertainty obtained in this study compared to the SH0ES result. The purple tilted lines indicate levels of discrepancy with the Planck measurement of the Hubble constant assuming a flat $\Lambda$CDM 
    cosmological model. The blue and red arrows and the corresponding annotations describe the main cause of the apparent decrease in the best-fit $H_{0}$ between the SH0ES measurements and model A of this work, and between models A B of this work. The new supernova modelling 
    developed in this study results in a decrease of the discrepancy between the SH0ES and Planck $H_{0}$ values by at least 30 per cent (model A) 
    and up to 50 per cent (model B).
    }
\label{H0values}
\end{figure}

We use a two-population model of type Ia supernovae to identify close analogs of the calibration supernovae in the Hubble flow, constrain the 
intrinsic and extrinsic effects related to peak magnitude-colour relation and derive the Hubble constant. The analogs are found in a probabilistic way as 
supernovae associated with high-stretch component of the stretch parameter distribution, which is well known to trace young and star forming 
supernova environments such as those in the calibration galaxies. We find a Milky Way-like mean extinction with $\widehat{R_{\rm B}}\approx 4$ 
for the discerned young supernova population. The result is consistent with the most likely expectations for interstellar extinction in the calibration galaxies as Milky-Way analogs, the extinction curve assumed for colour corrections in Cepheids from the SH0ES observations \citep{Riess2022} and recently obtained independent extinction measurement for NGC 5584, a SH0ES calibration galaxy \citep{Murakami2025}. On the other hand, it is discrepant with the dust model used in the Pantheon+ supernova compilation which assumes $\widehat{R_{\rm B}}\approx 3$ 
in the calibration galaxies with stellar masses $M_{\star}>10^{10}M_{\odot}$ (about 70 per cent of the calibration sample). Our modelling 
demonstrates that this discrepancy results in about $1.5$~km~s$^{-1}$~Mpc$^{-1}$ bias (30 per cent of the Hubble constant tension) in the Hubble constant 
derived from the the SH0ES Cepheid data and light curve parameters of type Ia supernovae from the Pantheon+ compilation (see Figure~\ref{H0values}).

We find tentative evidence for a higher mean host-galaxy reddening in the calibration galaxies than in the corresponding young population in the Hubble flow. 
The signal is primarily driven by a flatter distribution of reddened supernovae in the calibration sample than the young population in the Hubble flow. The combined effect of the revised extinction coefficient and the differential constraints on reddening leads to a reduction of the Hubble constant by about $2.8$~km~s$^{-1}$~Mpc$^{-1}$ 
(50 per cent of the Hubble constant tension) with respect to the SH0ES measurement (see Figure~\ref{H0values}). 

Possible differences between the mean reddening in the calibration sample and the corresponding population in the Hubble flow may occur due to a wide range of hidden 
variables relevant for predicting host-galaxy reddening from dust column densities (dust mass, dust distribution, supernova positions in their host galaxies etc.). The key problem is that there is no guarantee that these variables are sampled in the same way both in the calibration galaxies and the Hubble flow. Noticeable differences can arise merely as sampling errors. For example, with $0.5$~dex scatter in dust masses at fixed stellar mass \citep[typical scatter around observed the dust-to-stellar mass relation][]{Cortese2012}, we expect that the mean reddening 
in a calibration sample of equal size to the SH0ES one ($37$ host galaxies) can deviate from the corresponding mean measured for calibration analogs in the Hubble flow 
by $\delta\langle E(B-V)\rangle/\langle E(B-V)\rangle=\delta\langle M_{\rm dust}\rangle/\langle M_{\rm dust}\rangle=(\delta M_{\rm dust}/M_{\rm dust})/37^{1/2}\approx0.19$, 
solely due to insufficient sampling in the calibration sector (assuming conservatively the same distributions of stellar masses and other relevant variables).

Our analysis does not include corrections due to flux limits of the surveys included in the low-redshift ($z<0.15$) part of the Pantheon+ compilation. Accounting for these effects would most likely increase distance estimates at high redshifts and thus lower the derived Hubble constant. It would also 
improve population typing across redshift. Given the fact that survey selection effects are second-order correction in the Pantheon+ bias model, which is primarily driven by the dust and intrinsic emission model \citep[see Figure 1 of][]{Wojtak2024}, we expect possible effects on the Hubble constant estimation to be smaller than the bias corrections obtained in our study. We checked that fitting our models to supernovae in the redshift range restricted to $0.015<z<0.10$ (less affected by potential selection effects) results in a decrease of the derived 
Hubble constant values on average by $0.3$~km~s$^{-1}$~Mpc$^{-1}$. This motivates 
future more refined data analysis based on a full forward modelling combining survey 
selection effects and the two-population framework.

\begin{figure}
        \centering
        \includegraphics[width=\linewidth]{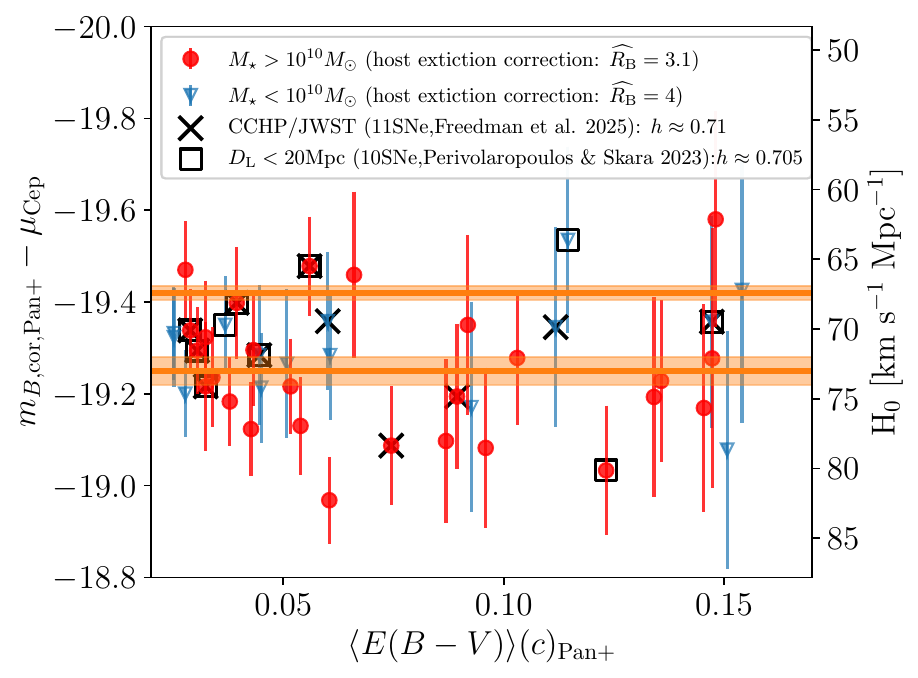} 
    \caption{Corrected peak magnitudes (as provided by the Pantheon+ data set) of 42 calibration supernovae relative to distance moduli from Cepheids, as a function 
    of the mean reddening $E(B-V)$ conditioned on supernova colour and based on the Pantheon+ dust model (see Appendix \ref{sec:appc}). 
    The filled circles (half-filled triangles) show 27 (15) supernovae for which Pantheon+ applies extinction correction 
    with $R_{\rm B}=3.1$ ($R_{\rm B}=4$). The plot shows also two samples of calibration galaxies from \citep{Freedman2024} and \citep{Perivolaropoulos2023} 
    which result in systematically lower best-fit values of the Hubble constant than the SH0ES result. It is apparent that both samples happen to minimise 
    the contribution from reddened supernovae or those with $R_{\rm B}=3.1$.
    }
\label{Pantheon_bias}
\end{figure}

Our extinction model in the young supernova population and thus in the calibration galaxies can be thought of as both observationally and theoretically 
motivated revision of the dust model assumed to estimate biases in the Pantheon+ supernova compilation. Figure~\ref{Pantheon_bias} demonstrates the sensitivity 
of the Hubble constant estimation to the Pantheon+ dust model. It is apparent that supernovae with extinction corrections based on $\widehat{R_{\rm B}}=3.1$ 
(27 out of 42 calibrators) pull the best-fit Hubble constant towards higher values, especially those highly reddened. These supernovae become intrinsically 
brighter in our modelling resulting in a decrease of the derived Hubble constant. A similar correction of the Hubble constant estimate was also obtained in our previous study 
based on direct replacement of the Pantheon+ dust model in the calibration galaxies with the Milky Way-like extinction \citep{Wojtak2024}. 
The bias caused by supernovae with underestimated extinction 
($R_{\rm B}=3.1$) in the Pantheon+ catalogue can be alternatively minimised by means of using only blue supernovae (minimally affected by extinction). 
\citet{Gall2024} showed that this results in $H_{0}\approx 70$~km~s$^{-1}$~Mpc$^{-1}$, which is consistent with the estimates obtained in our study.

Figure~\ref{Pantheon_bias} shows also the correspondence between our measurement of the Hubble constant and other analyses obtaining comparably 
low values of the Hubble constant. Reanalysis of the SH0ES and Pantheon+ data (including the original bias corrections and 
covariance matrix) by \citet{Perivolaropoulos2023} showed that type Ia supernovae at distances smaller than $20$ Mpc are about $0.08$~mag intrinsically 
brighter implying the best-fit Hubble constant at the mid-point between the SH0ES and Planck values ($H_{0}\approx 70.5$~km~s$^{-1}$~Mpc$^{-1}$). The obtained shift in the absolute magnitude is completely driven by the calibration supernovae (mean uncertainty of $0.06$~mag in distance moduli for 10 calibration supernovae relative to $0.4$~mag driven by peculiar velocities for 5 uncalibrated supernovae at comparable distances). Figure~\ref{Pantheon_bias} shows that the 10 calibration supernovae at distances $<20$~Mpc happen to be nearly free of the cases which are significantly reddened and found in massive ($M_{\star}>10^{10}M_{\odot}$) host-galaxies where the dust model employed in the Pantheon+ compilation underestimates 
extinction ($R_{\rm B}=3$). Selecting these supernovae mitigates effectively the impact of underestimated extinction and for this reason, it is not surprising 
that the resulting Hubble constant is consistent with the measurement obtained in our study.

Systematically lower best-fit value of the Hubble constant than SH0ES was also obtained by \citet{Freedman2024} using the James Webb Space Telescope observations 
of Cepheids in 11 SH0ES calibration galaxies within the distance range of the 14 closest ones. As shown in Figure~\ref{Pantheon_bias}, the selected calibration galaxies appear to have very little overlap with those where the Pantheon+ dust model assumes $R_{\rm B}\approx 3$. \citet{Riess2024} find a fair agreement between the JWST-based and the original 
SH0ES distance estimates for these galaxies, and suggest that the relatively low value of the Hubble constant is merely a sampling effect. Figure~\ref{Pantheon_bias} shows, however, that the selected and omitted supernovae do not constitute random subsamples of extinction corrections assumed in the two stellar-mass bins. This asymmetry matters because one of the extinction models is discrepant with $R_{\rm B}\approx 4$ found consistently in galaxies analogous to the calibrators, e.g. late-type galaxies \citep{Salim2018}, type Ia supernova host galaxies \citep{Rino-Silvestre2025} and the Milky Way \citep{Fitzpatrick2007,Schlafly2016}.

Our results demonstrate that a data-driven model of type Ia supernova standardisation with physically motivated extinction corrections is capable to explain 
at least 30 per cent and up to 50 per cent of the discrepancy between the SH0ES and Planck results. The lowest $H_{0}$ estimate obtained in our study is only 
$2.2\sigma$ different from the Planck cosmology. This tension level is sufficiently low to assume that the SH0ES+Pantheon+ and Planck data (or CMB data in general) 
are mutually consistent within the framework of a flat $\Lambda$CDM cosmology. 

The presented modelling of type Ia supernovae can be improved in 
several respects. Supernova populations defined by two peaks of the stretch distributions do not completely differentiate between 
young and old stellar environments \citep{Rigault2020,Ginolin2024a}. Modelling cross mixing between the related populations or adding extra information 
on host galaxies can help obtain a more accurate match between supernova environments in the calibration galaxies and the Hubble flow. 
Extra systematic effects may likely arise from Cepheid-based calibration. There is a range of analyses pointing to potential positive biases 
(overestimation of the Hubble constant) but none which would show the opposite effect. Systematically lower estimates of the Hubble constant relative 
to the SH0ES result are obtained by \citet{Freedman2024} from distance calibration methods that are independent of Cepheids but applied to the same calibration galaxies. 
Similar trends, although of smaller magnitude, result from alternative approaches to modelling Cepheid observations \citep[see e.g.][]{Hogras2025}. A wider range of possible shifts 
in the Hubble constant estimate with $\sigma_{H_{0}}\approx 1.7$~km~s$^{-1}$~Mpc$^{-1}$ can be also realised when one attempts to eliminate potential sources of 
systematics due to heterogeneity of observational data by excluding nearby Cepheids from the Local Group \citep{Kushnir2025}. These studies demonstrate that we cannot 
rule out that Cepheid observations may cause an extra bias in the $H_{0}$ estimate, of magnitude that is comparable to the effect shown in this study. Future tests 
should take into account a plausible option that systematics in the SH0ES measurement may involve a combination of the two factors rather than only one.

\begin{acknowledgements}
This work was supported by research grants (VIL16599,VIL54489) from VILLUM FONDEN. RW thanks Luca Izzo, Jo\~{a}o Duarte, Santiago Gonz\'{a}lez-Gait\'{a}n and Lucas Hallgren for helpful comments on the manuscript. The authors thank the anonymous referee for useful comments that helped improve this work.
\end{acknowledgements}

\bibliographystyle{aa}
\bibliography{master}

\onecolumn
\begin{appendix}
\section{Light curve parameters of type Ia supernovae in the Hubble flow}
\label{sec:appA}
\begin{figure*}
    \centering
    \begin{subfigure}[l]{0.45\textwidth}
        \centering
        \includegraphics[width=\linewidth]{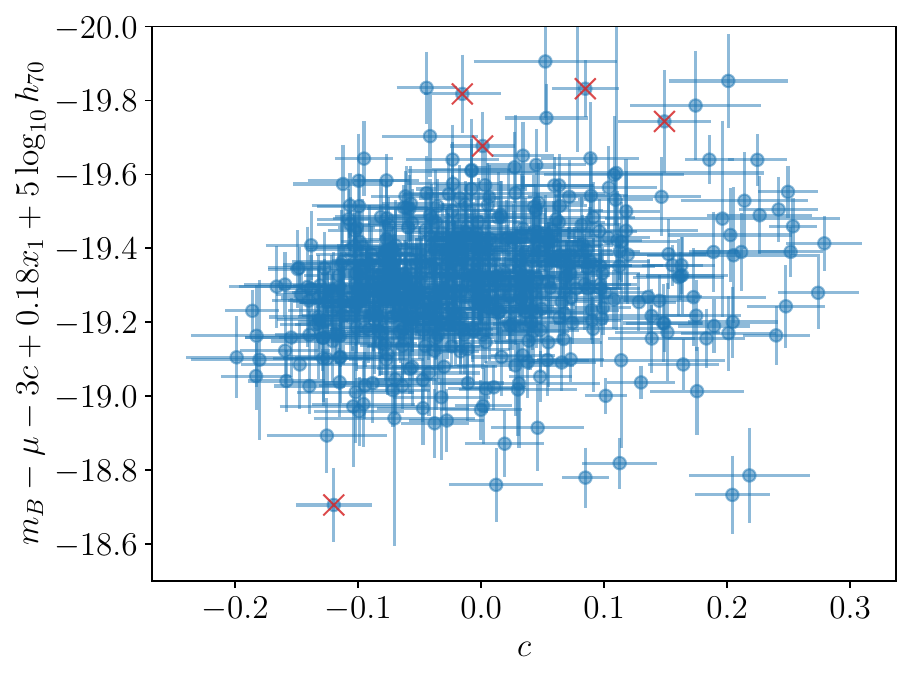} 
    \end{subfigure}
    \begin{subfigure}[r]{0.45\textwidth}
        \centering
        \includegraphics[width=\linewidth]{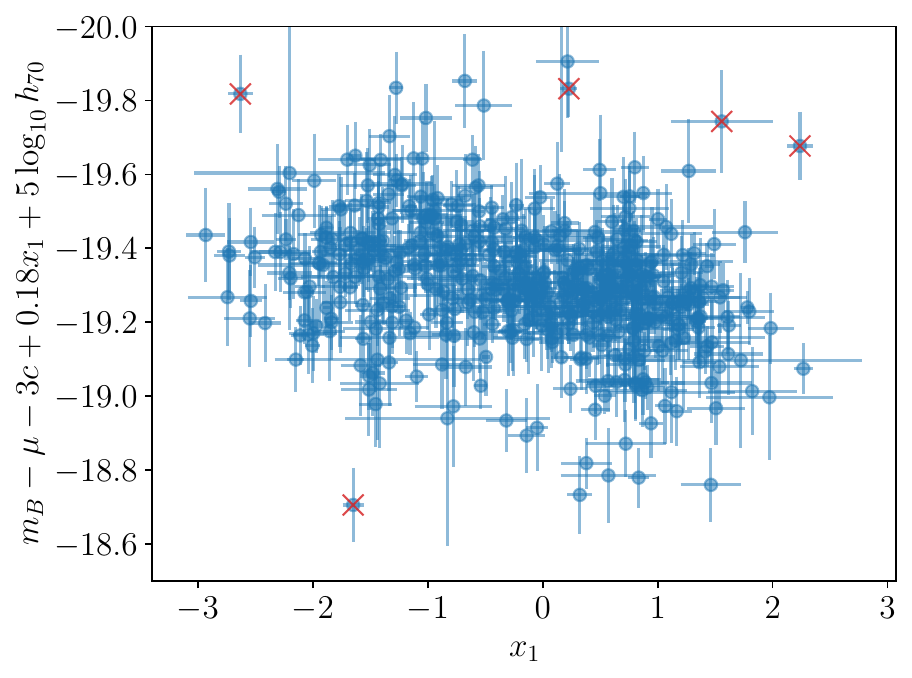} 
    \end{subfigure}
    \caption{Hubble residuals and light curve parameters of type Ia supernovae from the Hubble flow ($0.015<z<0.15$) of the Pantheon+ compilation. 
    Supernova absolute magnitudes shown on the vertical axes are computed for the standard linear corrections in stretch and colour. The five crosses mark 
    outlying supernovae which are omitted from the analysis.}
\label{data_HF_app}
\end{figure*}

Figure~\ref{data_HF_app} shows distance-corrected peak magnitudes of the selected supernovae as a function of the two remaining light curve parameters. 
The peak magnitudes are corrected for the first-order approximation of their dependence on $c$ and $x_{1}$ for better visualisation of the data.The correction 
-- the so-called Tripp calibration \citep{Tripp1998} -- involves linear terms in $c$ and $x_{1}$ so that
\begin{equation}
m_{\rm B}=M_{\rm B,T}+\mu(z)+\beta_{\rm T} c -\alpha_{\rm T} x_{1},
\label{eq:Tripp}
\end{equation}
where $M_{\rm B,T}$ is the supernova absolute magnitude. In this formulation, $M_{\rm B,T}$ is not a single-value parameter but a variable which includes 
all residuals (colour and stretch dependent) arising from the incompleteness of the Tripp calibration as a standardisation model. For the linear 
coefficient, we adopt $\beta_{\rm T}=3$ as a typical value found from fitting the Tripp calibration to cosmological samples of 
type Ia supernovae and $\alpha_{\rm T}=0.18$ obtained from the two-population modelling carried out in this study (see section~\ref{sec:results}). The errors 
in corrected magnitudes shown in Figure~\ref{data_HF_app} account for properly propagated uncertainties in $c$ and $x_{1}$, and their correlations. 
These errors are evaluated merely for the purpose of plotting the data and they are not used in our fits based on two-population model. 

\section{Model with high-stretch component of the old population}
\label{sec:appB}
\begin{figure*}
    \centering
    \begin{subfigure}[l]{0.45\textwidth}
        \centering
        \includegraphics[width=1.8\linewidth]{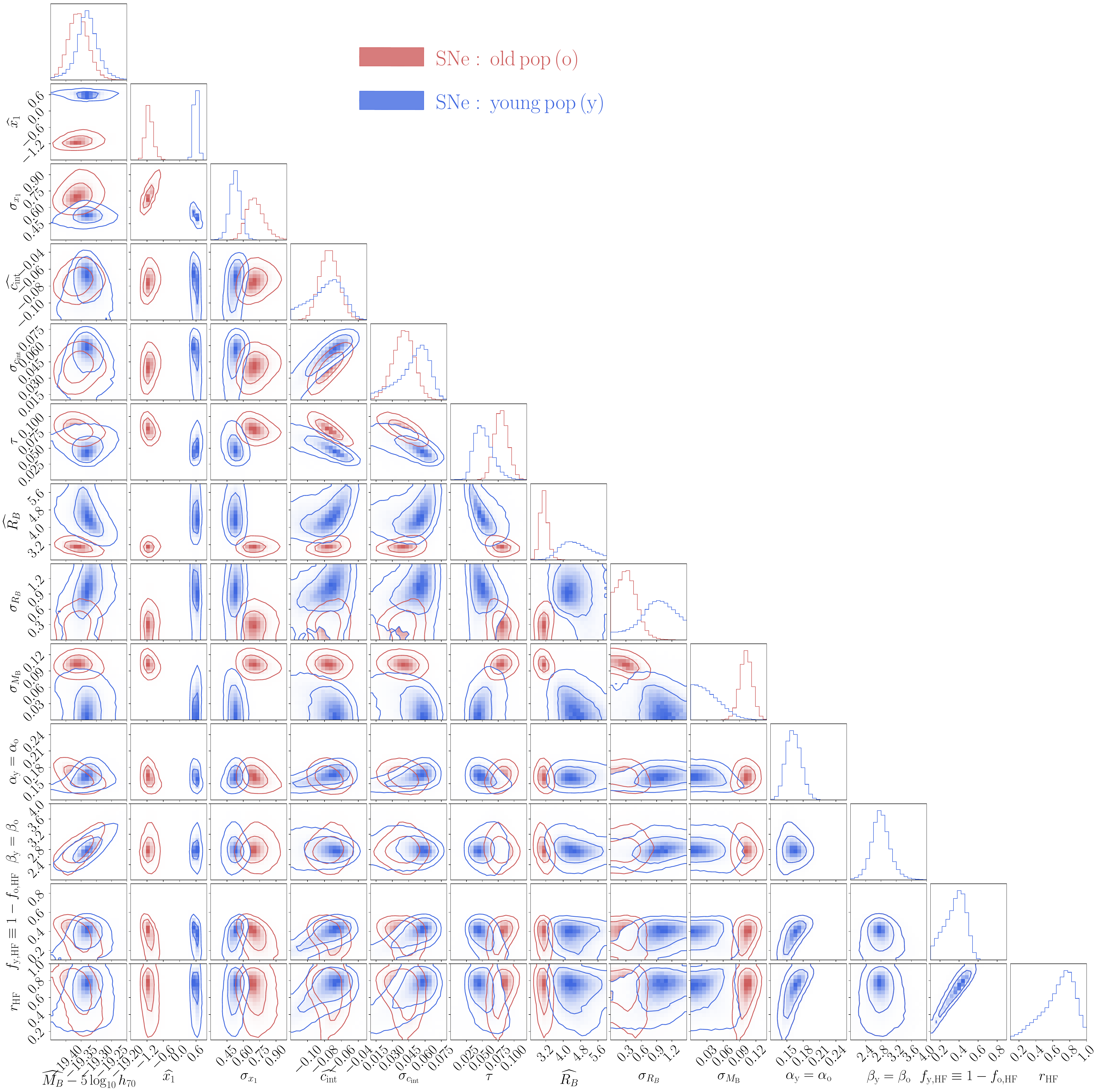} 
    \end{subfigure}
    \begin{subfigure}[t]{0.45\textwidth}
        \centering
        \includegraphics[width=0.6\linewidth]{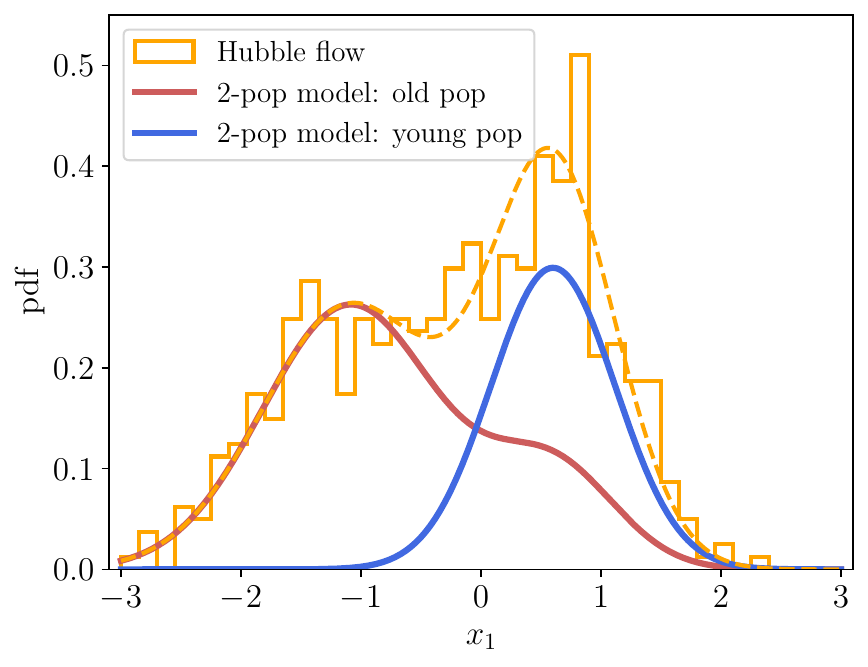} 
    \end{subfigure}
    \caption{Constraints on hyperparameters of the two-population mixture model from the analysis of type Ia supernovae in the Hubble flow from the Pantheon+ compilation. The red 
    and blue colours denote respectively the old and young populations, where the the low-stretch peak is fully populated by the old population and the high-stretch peak by a mixture of 
    both population. The contours show $1\sigma$ and $2\sigma$ credible regions containing 68 and 95 per cent 
    of two-dimensional marginalised probability distributions. The inset panel shows the stretch distributions of both supernova populations compared to the measurements.}
              \label{baseline_HF_ref}
    \end{figure*}

We model possible contribution of the old supernova population to the high-stretch peak by the following modification of the model probability given by eq.~\ref{prob_main}
\begin{equation}
p_{\rm prior}(\pmb{\phi})=f_{\rm{low-}X_{1}}p_{{\rm prior,low-}X_{1}}(X_{1})p_{\rm prior,o}(\pmb{\phi}|\pmb{\Theta_{\rm o}})+(1-f_{\rm{low-}X_{1}})p_{{\rm prior,high-}X_{1}}(X_{1})[rp_{\rm prior,y}(\pmb{\phi}|\pmb{\Theta_{\rm y}})+(1-r)p_{\rm prior,o}(\pmb{\phi}|\pmb{\Theta_{\rm o}})],
\label{prob_main_ref}
\end{equation}
where $r$ is an additional parameter equal to the fraction of the high-stretch component assigned to the young supernova population. The total fraction of the old and young supernova populations are given by $f_{\rm o}=f_{\rm{low-}X_{1}}+(1-f_{\rm{low-}X_{1}})(1-r)$ and $f_{\rm y}=(1-f_{\rm{low-}X_{1}})r$. The prior distributions of $X_{1}$ are directly shown in the above equation so that $X_{1}$ should be omitted from the vector of latent variable $\phi$. The results of fitting the model to the Hubble flow data are shown in Figure~\ref{baseline_HF_ref}. Table~\ref{bestmodels_HF_ref} summarises the fits based on the Hubble flow with or without the calibration sample.

\begin{table*}
\caption{Best-fit hyperparameters of the prior probability distributions of the modified two-population mixture model (see eq.~\ref{prob_main_ref}).}
\begin{center}
\begin{tabular}{lcccc}
\hline
 & \multicolumn{2}{c}{Hubble flow} & \multicolumn{2}{c}{Hubble flow + calibration sample} \\
 & old pop & young pop & old pop  & young pop  \\
 & $X_{1,{\rm low}}$ and $X_{1,{\rm high}}$ & $X_{1,{\rm high}}$ & $X_{1,{\rm low}}$ and $X_{1,{\rm high}}$ & $X_{1,{\rm high}}$ \\
\hline
 \\
$\widehat{M}_{\rm B}$ & $ -19.36 ^{+ 0.03 }_{- 0.03 }$ & $ -19.33 ^{+ 0.04 }_{- 0.03 }$ & $ -19.35 ^{+ 0.04 }_{- 0.04 }$ & $ -19.29 ^{+ 0.04 }_{- 0.04 }$ \\
& \\
$\widehat{c_{\rm int}}$ & $ -0.074 ^{+ 0.011 }_{- 0.011 }$ & $ -0.081 ^{+ 0.021 }_{- 0.024 }$ &   $ -0.080 ^{+ 0.010 }_{- 0.010 }$ & $ -0.062 ^{+ 0.011 }_{- 0.011 }$ \\
& \\
$\sigma_{c_{\rm int}}$ & $ 0.040 ^{+ 0.010 }_{- 0.010 }$ & $ 0.048 ^{+ 0.015 }_{- 0.017 }$ & $ 0.035 ^{+ 0.009 }_{- 0.009 }$ & $ 0.063 ^{+ 0.007 }_{- 0.007 }$ \\
& \\
$\tau$ & $ 0.080 ^{+ 0.011 }_{- 0.011 }$ & $ 0.052 ^{+ 0.014 }_{- 0.015 }$ & $0.085 ^{+ 0.010 }_{- 0.010 }$ & $ 0.044 ^{+ 0.010 }_{- 0.010 }$\\
& \\
$\widehat{R_{\rm B}}$ & $ 3.109 ^{+ 0.172 }_{- 0.173 }$ & $ 4.546 ^{+ 0.706 }_{- 0.651 }$ & $ 3.067 ^{+ 0.152 }_{- 0.155 }$ & $ 4.723 ^{+ 0.710 }_{- 0.671 }$ \\
& \\
$\sigma_{R_{\rm B}}$ & $ 0.319 ^{+ 0.189 }_{- 0.203 }$ & $ <1.08$ &  $ 0.303 ^{+ 0.178 }_{- 0.192 }$ & $ <1.11$ \\
& \\
$\sigma_{M_{\rm B}}$ & $ 0.102 ^{+ 0.011 }_{- 0.011 }$ & $ <0.042$ & $ 0.103 ^{+ 0.011 }_{- 0.011 }$ & $ <0.033$ \\
& \\
$\alpha$ & \multicolumn{2}{c}{$ 0.162 ^{+ 0.014 }_{- 0.014 }$} & \multicolumn{2}{c}{$ 0.177 ^{+ 0.012 }_{- 0.012 }$} \\
& \\
$\beta$ & \multicolumn{2}{c}{$ 2.795 ^{+ 0.233 }_{- 0.234 }$} & \multicolumn{2}{c}{$ 2.767 ^{+ 0.157 }_{- 0.158 }$} \\
& \\
$f_{\rm o,HF}=1-f_{\rm y,HF}$ & \multicolumn{2}{c}{$ 0.649 ^{+ 0.126 }_{- 0.113 }$} & \multicolumn{2}{c}{$ 0.555 ^{+ 0.064 }_{- 0.065 }$} \\
& \\
$r_{\rm HF}$ & \multicolumn{2}{c}{$0.662^{+0.189}_{-0.183}$} & \multicolumn{2}{c}{$ 0.805 ^{+ 0.100 }_{- 0.097 }$} \\
& \\
$f_{\rm o,cal}=1-f_{\rm y,cal}$ & \multicolumn{2}{c}{-} & \multicolumn{2}{c}{$\equiv 0$} \\
\end{tabular}
\label{bestmodels_HF_ref}
\end{center}
\tablefoot{The model allows the old supernova population to contribute to the high-stretch peak of the stretch distribution (see eq.~\ref{prob_main_ref}). 
The results were obtained from fitting type Ia supernovae in the Hubble flow (left) or from a joint analysis of the Hubble flow and calibration samples (right). Best-fit results are provided as the posterior mean values and errors given by credible intervals containing 68 per cent of the marginalised probabilities. Two parameters ($\alpha$ and $\beta$) are assumed to be the same in both supernova populations. 'HF' and 'cal' denote parameters used exclusively in the Hubble flow or the calibration sample.}
\end{table*}

\section{Mean reddening conditioned on supernova colour}
\label{sec:appc}
Using a model of supernova intrinsic colour and dust reddening one can find the expected distribution of reddening conditioned on the observed supernova colour. The resulting mean reddening is given by the following equation:
\begin{equation}
\langle E(B-V)\rangle (c) = \int (c-c_{\rm int})p_{\rm{prior},c_{\rm int}}(c_{\rm int})p_{\rm{prior},E(B-V)}(c-c_{\rm int})\textrm{d}c_{\rm int},
\end{equation}
where $p_{\rm{prior},c_{\rm int}}(c_{\rm int})$ and $p_{\rm{prior},E(B-V)}(E(B-V))$ are the prior distributions of intrinsic colours and reddening with hyperparameters measured from supernova data. We note that for $c\lesssim \widehat{c_{\rm int}}$ the 
reddening probability peaks at $E(B-V)=0$ but the mean reddening remains positive.

\end{appendix}

\end{document}